\documentclass[fleqn,usenatbib]{mnras}
\usepackage{newtxtext,newtxmath}
\usepackage[T1]{fontenc}
\usepackage{ae,aecompl}
\usepackage{enumerate}
\usepackage[switch, modulo]{lineno}
\usepackage[toc,page]{appendix}

\usepackage{graphicx}	
\usepackage{amsmath}	
\usepackage{amssymb}	

\newcommand{\LC}[1]{\textcolor{black}{#1}}

\title[Background light prediction on astronomical images]{The PAU Survey: Background light estimation with deep learning techniques}

\author[Cabayol-Garcia et al.]{L.~Cabayol-Garcia$^{1}$\thanks{E-mail:lcabayol@ifae.es},
M.~Eriksen$^{1}$\thanks{E-mail: eriksen@pic.es} \thanks{Also at Port d'Informaci\'{o} Cient\'{i}fica (PIC), Campus UAB, C. Albareda s/n, 08193 Bellaterra (Cerdanyola del Vall\`{e}s), Spain},
A.~Alarc\'on$^{2,3}$,
A.~Amara$^{4}$,
J.~Carretero$^{1}$\footnotemark[3],
\newauthor
R.~Casas$^{2,3}$, 
F.~J.~Castander$^{2,3}$,
E.~Fern\'andez$^{1}$, 
J.~Garc\'ia-Bellido$^{5}$,
\newauthor
E.~Gaztanaga$^{2,3}$,
H.~Hoekstra$^{6}$, 
R.~Miquel$^{1,7}$, 
C.~Neissner$^{1}$\footnotemark[3],
C.~Padilla$^{1}$
\newauthor
E.~S\'anchez$^{8}$,
S.~Serrano$^{2}$,
I.~Sevilla-Noarbe$^{2}$, 
M.~Siudek$^{1}$,
P.~Tallada$^{8}$\footnotemark[3],
L.~Tortorelli$^{9}$
\\ \\
$^{1}$Institut de F\'{\i}sica d'Altes Energies (IFAE), The Barcelona Institute of Science and Technology, 08193 Bellaterra (Barcelona), Spain \\
$^{2}$Institute of Space Sciences (ICE, CSIC),  Campus UAB, Carrer de Can Magrans, s/n,  08193 Barcelona, Spain\\
$^{3}$Institut d'Estudis Espacials de Catalunya (IEEC), 08193 Barcelona, Spain\\
$^{4}$Institute of Cosmology \& Gravitation, University of Portsmouth, Dennis Sciama Building, Burnaby Road, Portsmouth PO1 3FX, UK\\
$^{5}$Instituto de Fisica Teorica UAM/CSIC, Universidad Autonoma de Madrid, 28049 Madrid, Spain\\
$^{6}$Leiden Observatory, Leiden University, Leiden, The Netherlands\\
$^{7}$Instituci\'o Catalana de Recerca i Estudis Avan\c{c}ats, E-08010 Barcelona, Spain\\
$^{8}$Centro de Investigaciones Energ\'eticas, Medioambientales y Tecnol\'ogicas (CIEMAT), Madrid, Spain\\
$^{9}$Institute for Particle Physics and Astrophysics, ETH Z\"urich, Wolfgang-Pauli-Str. 27, 8093 Z\"urich, Switzerland\\
}

\date{Accepted XXX. Received YYY; in original form ZZZ}

\pubyear{2019}

\begin{document}
\label{firstpage}
\pagerange{\pageref{firstpage}--\pageref{lastpage}}
\maketitle
 
\begin{abstract}

\LC{In any imaging survey, measuring accurately the astronomical background light is crucial to obtain good photometry. This paper introduces \textsc{BKGnet}, a deep neural network to predict the background and its associated error. \textsc{BKGnet} has been developed for
data from the Physics of the Accelerating Universe Survey (PAUS), an imaging survey using a 40 narrow-band filter camera (PAUCam).} Images obtained with PAUCam are affected by scattered light: an optical effect consisting of light multiply reflected that deposits energy in specific detector regions contaminating the science measurements. Fortunately, scattered light is not a random effect, but it can be predicted and corrected for.  We have found that \textsc{BKGnet} background predictions are very robust to distorting effects, while still being statistically accurate. On average, the use of BKGnet improves the photometric flux measurements by $7\%$ and up to $20\%$ at the bright end. \textsc{BKGnet} also removes a systematic trend in the background error estimation with magnitude in the $i$-band that is present with the current PAU data management method. With \textsc{BKGnet}, we reduce the photometric redshift outlier rate by $35\%$ for the best $20\%$ galaxies selected with a photometric quality parameter. 
 \end{abstract}




\section{Introduction}

\LC{The positions, fluxes and other properties of galaxies and stars can be determined by analysing images of the sky. Modern imaging surveys can cover large areas of sky efficiently, resulting in measurements for large numbers of galaxies to faint magnitudes (e.g. DES \citep{DESDR1}). Improving the accuracy of the measurements is crucial for future weak lensing surveys, e.g. LSST and {\it Euclid} \citep{LSST, Euclid}, to ensure that the results are not dominated by systematic errors.}

\LC{For imaging surveys, accurate flux measurements are
essential: they are used to select samples of galaxies, or to infer their physical properties. A key step towards  
a reliable flux estimate is the determination of the background, which needs to be subtracted. The main source of background is the brightness of the night sky, which may vary due to a range of effects, such as illumination by the Moon, airglow and light pollution. Instrumental effects can contribute as well, and in this paper we focus on scattered light, which is the result of light deflecting from the instrument optical path appearing at a different region of the detector \citep{photometrybook}.} \\

\LC{Different approaches have been used to estimate the sky background \citep{Bijaoui,Newell}, and  example implementations include \textsc{DAOPHOT} \citep{DAOPHOT} and \textsc{SExtractor} \citep{SExtractor}. \textsc{DAOPHOT} measures the background as the mode of the uniformly scattered pixels at a certain Full Width at Half Maximum (FWHM) of the given target source. On the other hand, \textsc{SExtractor} meshes the background and reconstructs a 'background map' with the background estimated at each particular mesh location. Other methods aim to be more robust in the presence of nearby sources. In \citet{Teeninga}  the background is estimated at a location without nearby sources, while \citet{Popowicz}, is based on the removal of small objects and an interpolation of missing pixels. In this paper we propose a new approach based on a deep neural network to predict the background and its associated error.} \\

\LC{Over the last few years, deep learning algorithms have resulted in revolutionary advances in machine learning and computer vision \citep{computervision}. Theoretical breakthroughs in training deep Neural Networks (NN) \citep{werbos},  or Convolutional Neural Networks (CNN) \citep{lecun89,lecun98,cnn_understanding} together with powerful and efficient parallel computing provided by Graphics Processing Units (GPUs) \citep{Alexnet} have lead to  groundbreaking improvements across a variety of applications. The number of deep learning projects in cosmology is quickly increasing. This includes astronomical object classification \citep{objseq_class, Cabayol}, \citep{Cabayol},  gravitational wave detection \citep{GW} and directly constraining cosmological parameters from mass maps \citep{DL_mapsConstrains, ETH2}.} \\

\LC{Extracting the source photometry requires a significant amount of data engineering and parameter tweaking. This can be particularly challenging for noisy sources. Deep learning has already been successfully applied to different steps in source photometry extraction. Examples include point source detection \citep{DeepSource}, cosmic ray detection \citep{deepCR} or Point Spread Function (PSF) modelling \citep{psf}. Deep learning has also been used to directly estimate photometric redshifts from images \citep{photoz, photoz_img}. These algorithms implicitly include steps for the background subtraction. Understanding these image processing steps can optimize the performance of e.g. redshift estimation and galaxy classification.}\\

\LC{Our goal is to develop and test a deep learning background subtraction method using data from the Physics of the Accelerating Universe Survey (PAUS). PAUS is an imaging survey that measures high precision photo-$z$s to faint magnitudes ($i_{\rm AB} < 22.5$), while covering a large area of sky \citep{photoz-Marti}. This is possible thanks to the PAUCam instrument \citep{PAUCAM_Francisco, PAUCam-Padilla,PAUCam}, an optical camera equipped with 40 narrow bands (NB) covering a wavelength range from 450nm to 850nm \citep{Casas2016}. 
PAUS reaches a photo-$z$ precision $\sigma(z)/(1+z) \sim 0.0035$ for the best 50\% of the sample, compared to typical precision of 0.05 for broad band measurements  \citep{photoz-Martin}. The scientific goals of PAUS include the measurement of intrinsic alignments of galaxies out to $z \sim 0.75$, the study of their
spectral energy distributions (SEDs), detailed studies of intermediate-scale cosmic structure \citep{Stothert}, and improvements of image simulations \citep{Tortorelli2018}}.\\

 \LC{PAUS imaged the COSMOS field as a calibration area given the availability of spectroscopic redshifts. The PAUS photo-$z$ catalogue for the full COSMOS sample with $i_{\rm AB} <22.5$ contains outliers when compared to the spectroscopic redshifts. Some of these outliers simply arise from noisy photometry, but others are due to a strongly varying continuum produced by scattered light. The excess scattered light decreases the signal-to-noise ratio (SNR) of the photometric measurement and also alters the statistics of the values of the pixels from which the continuum is estimated. These effects can potentially bias the flux measurements.} \\
 
In this paper, we present \textsc{BKGnet}, a convolutional neural network capable of learning the underlying behaviour of scattered light and other distorting effects present in the PAUCam images.  \textsc{BKGnet} predicts the background at the location of the target sources and the error associated with the background prediction. Although it is built to improve the PAUS photometry, \textsc{BKGnet} can be also be applied to other future imaging surveys such as LSST \citep{LSST} and \textit{Euclid} \citep{Euclid}. The code is available at \url{https://gitlab.pic.es/pau/bkgnet}. \\

The structure of this paper is as follows. In section 2, we  describe the PAU Survey and the PAUCam camera and present the modelling of scattered light using scattered-light templates.  In section 3, we introduce the specific network we have developed, as well as defining the training and testing process. Sections 4 and 5 contain the results obtained for simulated and real PAUCam images, respectively. In section 6, we validate the network predictions on real target locations and we conclude and summarise in section 7.

\section{Modelling scattered light}
\label{sec:SL_modelling}

PAUCam images contain substantial amounts of scattered light, which mostly affect the edge regions of some CCDs. Scattered light increases the amount of background in the affected regions and distorts the expected statistics of the pixel values used to estimate the background. Therefore, the scattered light present in PAUCam images can lead to an incorrect estimate of the background if not properly modelled, thus biasing the photometry. Moreover the elevated background lowers the SNR of the measurements.
In 2016 the PAU camera was modified in order to mitigate the effect of scattered light by introducing baffles on all the edges of the NB filters of each filter tray. Although this reduced the amount of scattered light, residuals still remain. In the latest COSMOS data reduction, around 8\% of exposures taken before the camera intervention are flagged as affected by scattered light, and therefore dismissed. After the intervention, this number reduced to 5\% of the exposures, such that on average $7\%$ of data in the COSMOS field are lost due to scattered light.
In this section, we present the PAUCam scattered light model we are using throughout the paper.

\subsection{The PAUS observations}
\label{s2:SL&PAU}
 PAUS has been observing since the 2015B semester and as of 2019A, PAUS has taken data for 160 nights. The current data covers 10 deg$^2$ of the CFHTLS fields\footnote{http://www.cfht.hawaii.edu/Science/CFHTLS\_Y\_WIRCam\\/cfhtlsdeepwidefields.html} W1, W2; 20 deg$^2$ in W3 and 2 deg$^2$ of the COSMOS field\footnote{http://cosmos.astro.caltech.edu/}. The PAUS data are stored at the Port d'Informaci\'o Cient\'ifica (PIC), where the data are processed and distributed \citep{Tonello}.\\
 
 In this paper we focus only on the data from the COSMOS field, which were taken in the semesters 2015B, 2016A, 2016B and 2017B (the low efficiency was caused by bad weather). The COSMOS field observations comprise a total of 9749 images, 343 images for each NB. From these images, 4928 were taken before the camera intervention and 4821 after. The basic exposure times in the COSMOS field are 70, 80, 90, 110 and 130 seconds from the bluest to the reddest filter trays.

\subsection{PAUS images processing}

\LC{The PAUCam instrument \citep{PAUCAM_Francisco, PAUCam-Padilla,PAUCam} is an optical camera equipped with 40 narrow bands (NB), covering a wavelength range from 450nm to 850nm \citep{Casas2016}. The NB filters have 13nm FWHM and a separation between consecutive bands of 10nm. The camera is also equipped with $ugrizY$ broad band filters that so far have been mainly used by external observers. The camera has 18 red-sensitive fully depleted Hamamatsu CCD detectors \citep{Casas2012}, although only the 8 central CCDs are used for NB imaging. Each CCD has 4096x2048 pixels with a pixel scale of 0.26 arcsec/pix. The NB filter set effectively measures a high resolution photometric spectrum ($R\approx 50$).}\\

\LC{In this project we use images that have already been corrected for various instrumental effects \citep{PAUimage} in the PAUS \textsc{nightly} pipeline. This pipeline performs basic instrumental de-trending processing. The electronic effects are corrected using a master bias, which is an observation with the shutter closed and zero exposure time. To correct pixel-to-pixel variations we use dome flats, which are obtained by imaging a uniformly illuminated screen. The  astrometry of the narrow band images is calibrated comparing to the positions of \textit{GAIA} DR2 stars \citep{GaiaDR2}. The photometry calibration is done relative to SDSS stars \citep{PAUcalib}.} \\

Once the images have been corrected, we perform forced photometry to extract the galaxy flux. The current \textsc{PAUdm} pipeline, similarly to \textsc{DAOPHOT}, predicts the background noise as the median of the pixels within a ring placed around the target source. However, this algorithm requires a (fairly) flat background for an accurate estimate. This assumption breaks down when either the annulus or source extraction regions are affected by scattered light. In addition, other sources of errors in the background estimation are undetected sources, cosmic rays or cross-talk. In order to minimize the effect of any of these artifacts, the pixels inside the annulus are  3-$\sigma$ clipped before computing the median. \\

\LC{The default \textsc{PAUdm} radii for the annulus region are $r_{\rm in} = 30$ and $r_{\rm out} = 45$ pixels \citep{PAUimage}. The annulus is selected to be sufficiently far away from the galaxy to avoid light leaking inside the ring and not too far so that the background is representative. Throughout this paper we use the default configuration to compare this commonly used approach to our deep learning algorithm. However, in Appendix~\ref{A2} we study the effect of a variable annulus.}
\subsection{Scattered-light templates}
\label{s3:skyFlats}

\begin{figure*}
\includegraphics[width=1\textwidth]{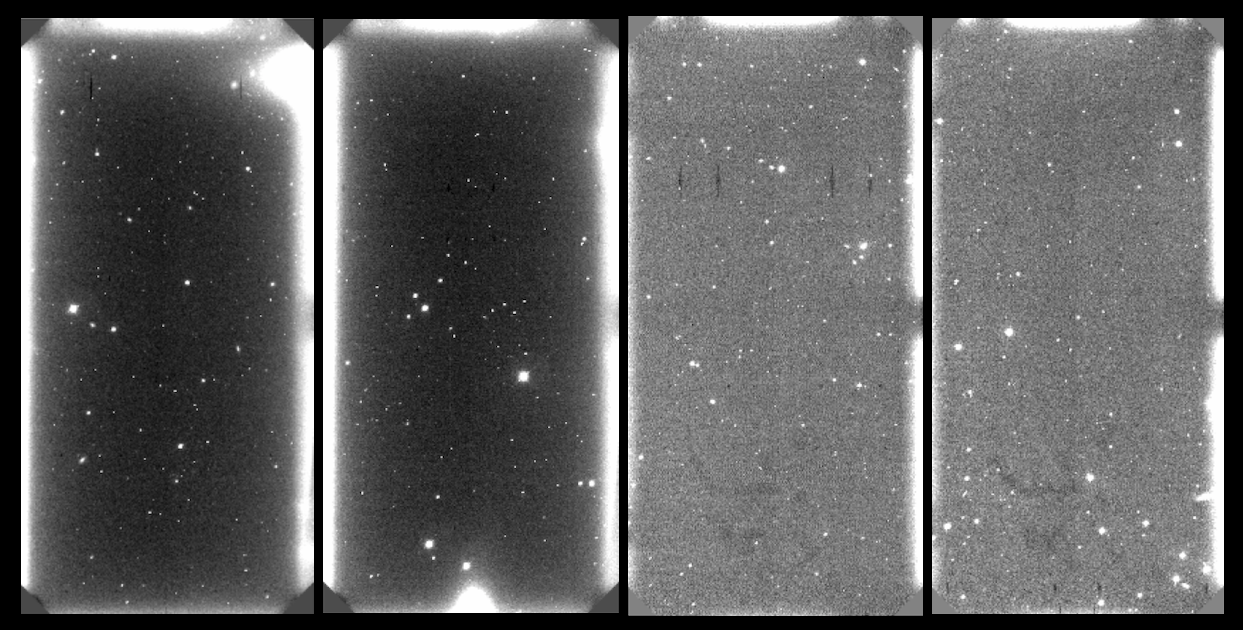}
\centering
\caption{Images taken with the PAUCam, corresponding to the NB685 filter. \emph{Left:} The first two images correspond to PAUCam images before the camera intervention. Notice that both exhibit the same scattered light pattern. \emph{Right:} The two images on the right correspond to PAUCam images after the intervention. Again, both present the same scattered light pattern, but different to the first two images on the left. This shows the changes in scattered light patterns with the intervention.}
\label{Fig1:scatteredLight_pattern}
\end{figure*}

\begin{figure}
\includegraphics[width=0.5\textwidth]{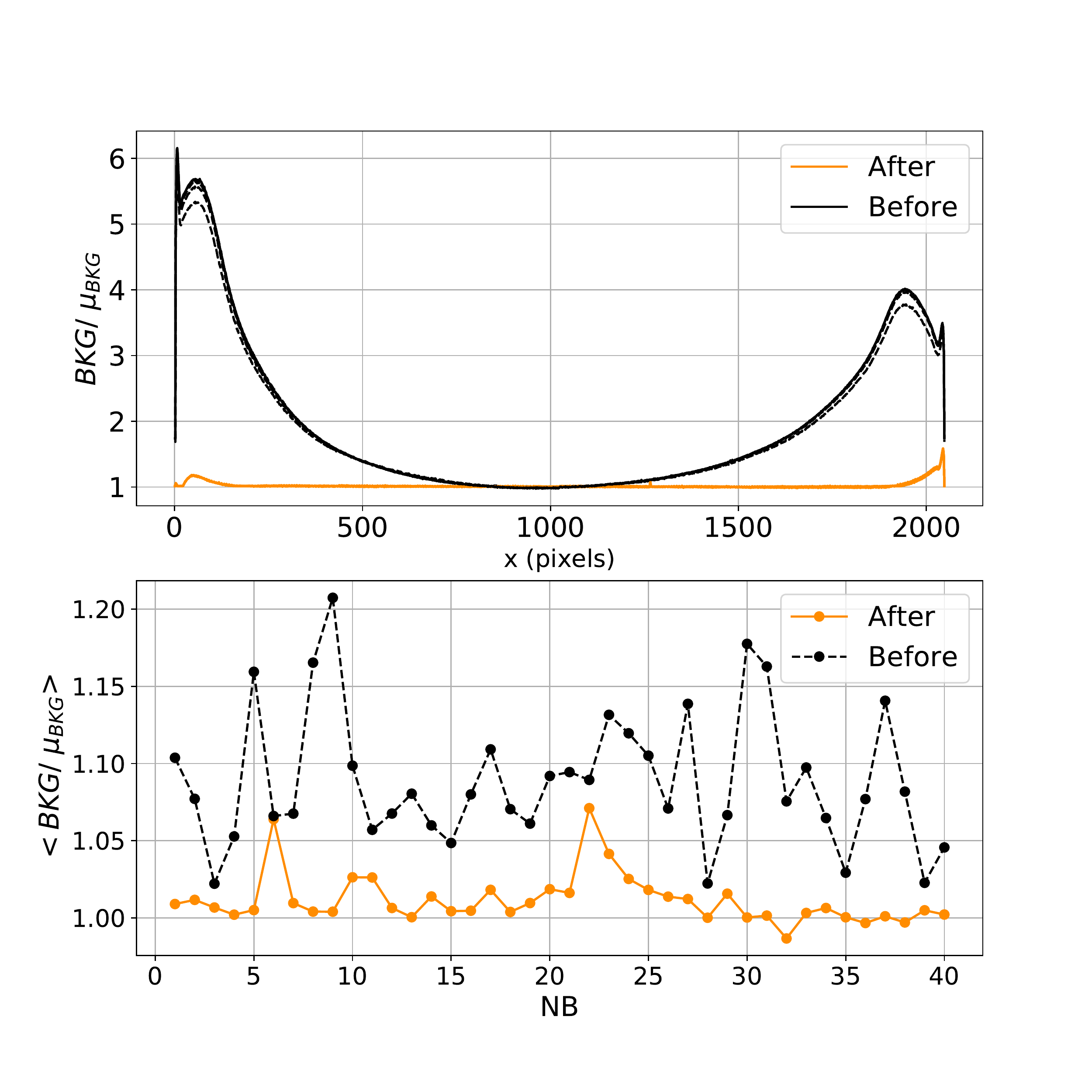}
\centering
\caption{\emph{Top}: Normalised background light content in each pixel as a function of the pixel position in the image for different images before (black dashed line) and after (orange solid line) the camera intervention. Each pixel value is divided by the mean background in the image. Regions without scattered light should fluctuate around unity. Regions affected by scattered light should be above unity. \emph{Bottom}: Mean value of the normalised background curves considering all the images taken in that band, for the 40 narrow photometric bands.}
\label{Fig2:intervention}
\end{figure}

Figure \ref{Fig1:scatteredLight_pattern} shows four PAUCam images in the NB filter NB685 before the camera intervention (first and second images on the left) and after the camera intervention (third and fourth images). They show scattered light near the edges of the CCD, displaying a spatially varying amount of scattered light. The scattered light patterns change from before the intervention (two images on the left) to after (two images on the right). The scattered light pattern is also filter dependent. The images taken in each filter show their own  distinctive patterns, meaning that the pattern depends on the filter used.\\

One way to quantify and model the scattered light is to create background pixel maps per NB. This is done with the following steps:
\begin{description}
\item[i. \emph{Select images}:] Select a group of NB images from the same bands, since they have the same scattered light pattern.
\item[ii. \emph{Compute median}:] For each of the images, compute the median background level in the central regions, $\mu_{\rm BKG}$, which are unaffected by scattered light.
\item[iii. \emph{Estimate ratios}:] Divide every image by its median to  obtain a pixel ratio map.
\item[iv. \emph{Mask sources}:] Mask the images sources by masking all pixels above a given pixel ratio threshold.
\item[v. \emph{Combine images}:] Combine all individual pixels maps with a median to get a single scattered-light template (SLT) for all the selected images.
\end{description}

If the background were flat and followed Poisson statistics, all pixels in the ratio map should fluctuate along unity. However, if the image is affected by scattered light, the scattered-light templates in affected regions will have a value above unity. We can understand this ratio as approximately the percentage of extra light (scattered light) compared to the flat background.  Notice that this model takes into account that scattered light depends on the amount of light falling on the CCD. \\

The procedure in step (v) can be written as
\begin{equation}
{\rm SLT}(x,y) =  \text{median}_j\Big[\frac{ I_j(x,y)}{\mu_{\rm BKG}}\Big],
\label{eq1:skyFlat} 
\end{equation} where $I_j$ is image $j$ and the median is over the selected images (step i). To determine the amount of scattered light we can follow the previous procedure to step [iii]. This way we obtain normalised background images that should fluctuate around unity if they contain a flat background, but would have values above one if they are affected by scattered light.\\

The Figure \ref{Fig2:intervention} top panel shows some of these normalised images for the NB685 filter. It shows the background pixel value from side to side of the image before (black dashed line) and one after (orange solid line) the camera intervention. The plot shows an increasing background on the edges of the CCD before the camera intervention. After the intervention, the amount of scattered light is considerably reduced. Unfortunately it is still present and thus needs to be accounted for. \\

We can use all the normalised background images in a given NB to create a general scattered-light template for that band (also splitting before/after the intervention). The bottom panel in Figure \ref{Fig2:intervention} shows the resulting mean of each scattered-light templates (one per band) as a function of NB. The mean of the scattered-light templates gives information about the amount of scattered light in a given band. We can clearly see the effect of the intervention on the amount of scattered light, which is reduced.

\subsection{Scattered-light templates as scattered light correcting method}
If the scattered-light templates modelling is sufficiently accurate, it can be used to correct the scattered light on PAUCam images. Assuming that all the images from a given NB follow the same scattered light pattern scaled by the CCD sky background, a way of correcting scattered light would be 
\begin{equation}
 \tilde{I}(x,y) = I(x,y) - (\text{SLT}(x,y) - 1)  \mu_{\rm BKG},
\label{eq2:skyFlat_correction}
\end{equation} where we subtract from a given target image ($I(x,y)$) the scattered-light templates scaled by the mean background of such image ($\mu_{\rm BKG}$).
Notice that instead of subtracting the scattered-light templates, we subtract the scattered-light templates without the flat sky background. This way, the regions without scattered light are barely affected.  \\
 
\begin{figure*}
\includegraphics[width = 1\textwidth]{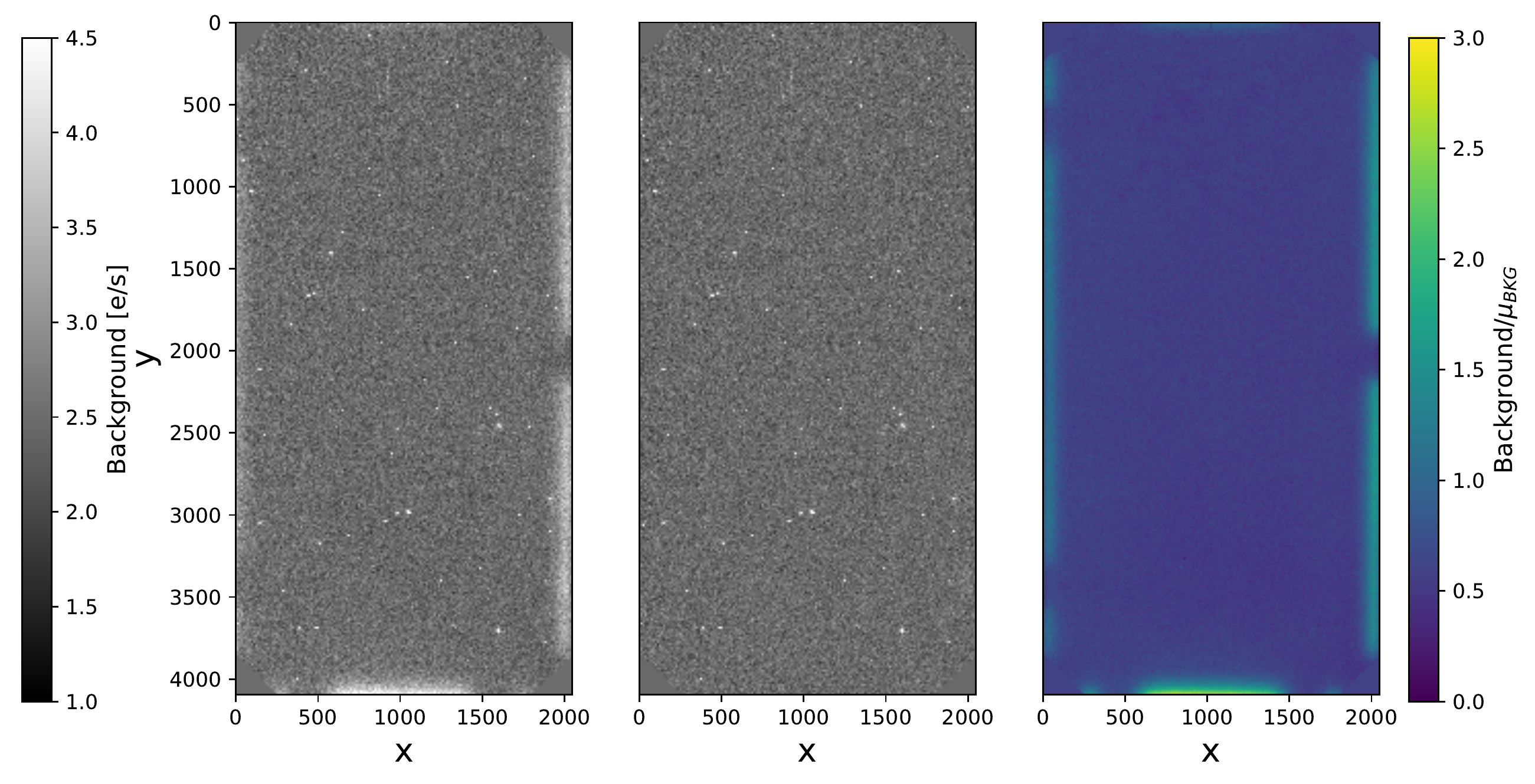}
\centering
\caption{\emph{Left:} Image taken in the NB685 filter showing a scattered light pattern on the edges. \emph{Middle:} Previous image corrected with the scattered light template. \emph{Right:}  The scattered-light template generated with equation \ref{eq1:skyFlat} considering all images taken the same observation night as the original image.}
\label{fig3:corrected_image}
\end{figure*}

Figure \ref{fig3:corrected_image} shows the original CCD image (left), after correcting with the scattered-light templates (middle) and the scattered-light templates used for correction (right). Visually, the scattered light pattern in the original image (left) disappears after applying the scattered-light templates correction (middle). However, although the correction seems visually almost perfect, this method has a drawback. Even though scattered light follows approximately a pattern given a band, there might be fluctuations due to other external conditions. For example, the weather, Moon illumination and other observing conditions may induce variations between different observations in a NB. To be more precise estimating the correction, one should create a template per band and per night, such that the observing conditions are similar. However, for creating a scattered-light templates per night, there might be a insufficient number of images to obtain an accurate modelling of the correction pattern. Bright stars also contribute to scattered light and this cannot be corrected with the scattered-light templates.\\

\begin{figure}
\includegraphics[width=0.5\textwidth]{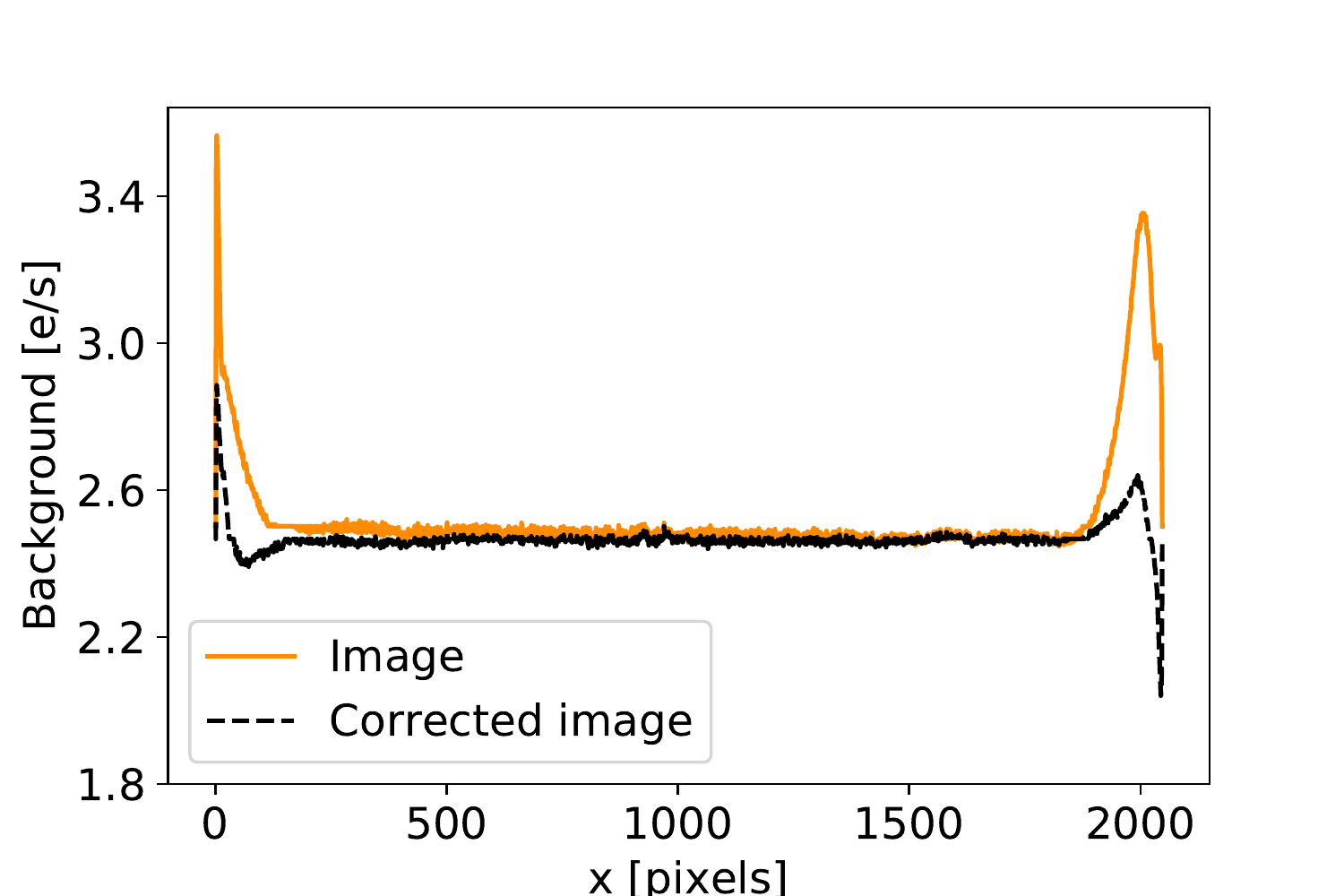}
\centering
\caption{Background pixel values across the image. The original image (orange solid line) displays high peaks on the edges caused by scattered light. After correcting with the scattered-light templates (dashed black line) the peaks are reduced, but some residuals remain. The images are in e/s.}
\label{fig4:badCorrection}
\end{figure}

Figure  \ref{fig4:badCorrection} shows the background level for a specific image in the NB685 filter before and after the correction with the scattered-light templates. In this case, the image is corrected without considering any split on night to generate the scattered-light templates. This means that all images, despite being observed on different night and with different observing conditions are used to build the scattered-light templates. The image without correction displays large peaks at both edges and those are clearly corrected by the scattered-light templates. However, both sides of the CCD still show bumps that are caused by scattered light residuals.

\section{\textsc{BKGnet}: A Deep Learning based method to predict the background}
\label{sec:BKGnet}
In this section we start by describing the \textsc{BKGnet} architecture. We then describe our training and test samples and describe the training process. As a reference, in Appendix \ref{A1} we introduce the basics of deep learning and convolutional neural networks, together with some terminology definitions.

\subsection{Neural network architecture}
 \textsc{BKGnet}\footnote{https://gitlab.pic.es/pau/bkgnet} is built using the \textsc{PyTorch} library \citep{pytorch}. It has two main blocks: a convolutional neural network (CNN) and a linear neural network.\\

\begin{figure*}
\includegraphics[width=0.9\textwidth]{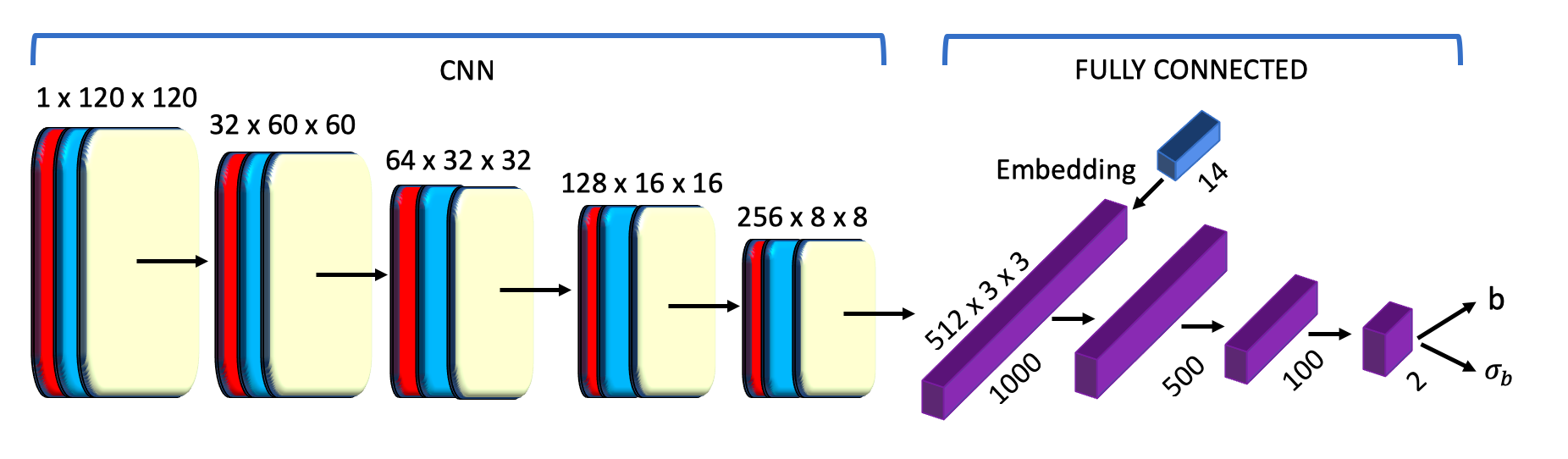}
\centering
\caption{BKGnet scheme: The first set of layers corresponds to a Convolutional Neural Network to which one inputs the images. The CNN output, together with extra information are input to a linear neural network. The numbers on each of the convolutional layers represent the layer's dimension. The first number corresponds to the number of channels. The second and third numbers are the dimension of the stamp in that layer. }
\label{fig6:bkgNet_scheme}
\end{figure*}

Figure \ref{fig6:bkgNet_scheme} shows the \textsc{BKGnet} architecture.  The CNN block handles the information coming from the image itself, as the background we want to recover is encoded in the pixel values. The inputs are 120x120 pixels stamps containing the target galaxy in the center. \LC{This choice for the stamp size is a compromise between having enough pixels whilst keeping the computing requirements (memory, GPU) within reasonable limits.}\\

\LC{As Figure \ref{fig6:bkgNet_scheme} shows, the CNN contains 5 blocks of convolutional layer (red layer), pooling layer (yellow layer) and batch normalization layer (blue layer). In each convolutional layer, the network learns to gradually capture different features in the image. The first layers learn low-level features, like edge detection, while having more layers 
leads the network to learn high-level features \citep{CNN_vis}.}  \\

\LC{The scattered light model depends on parameters that are not encoded in the stamps. These are the position of the stamp in the original image, the NB used to observe the galaxy and a before/after intervention flag informing the network when the galaxy was observed. We also include the target galaxy magnitude from a reference catalogue, as it contains  information about the number of pixels that are affected by the galaxy. To help the network learning the scattered light patterns, the previously mentioned parameters are provided to the linear neural network, together with the CNN's output. The magnitude of the galaxy and the coordinates of the image are added as fixed parameters per stamp.} \\

\LC{The NB filter and the intervention flag are discrete variables with forty possible values for the band (1-40) and two for the intervention flag (0/1). The combination of the two is, however, directly related to the scattered light pattern, and we can effectively convert these two discrete variables into a single one with values from 1 to 80. We add the band and intervention information using an embedding. The embedding replaces each combination with 10 trainable parameters. Before embedding the band and intervention information, \textsc{BKGnet} learns to encode each scattered light pattern using the ten numbers that should best characterise the pattern. Therefore, the linear network receives the output from the convolution layers, two galaxy coordinates, the magnitude and ten numbers from the embedding. }

\subsection{Data: training and test samples}
\label{sec:data}
\textsc{BKGnet}'s inputs are stamps with the target galaxy in the center. However, to train the network we use empty CCD positions, meaning regions where there are not target sources. This way, we can estimate the ground truth background value at the central CCD region (where there is supposed to be a target galaxy) and train the network to recover this value. The estimation of the true background values used as training sample labels is done by computing the mean background inside a circular aperture of a given fixed radius in the central region of the stamp. Therefore, these measurements have an associated uncertainty that directly depends on the aperture radius. Assuming that the background is purely Poissonian, then
\begin{equation}
    \sigma^2_{\rm label} = \frac{N_{\rm a} b}{t_{\rm exp}},
    \label{laberr}
\end{equation} where $t_{\rm exp}$ is the exposure time, $b$ is the background estimated as the mean of the pixels inside the aperture, i.e. the background label, and $N_{\rm a}$ is the number of pixels inside the circular aperture, directly related with the choice of aperture radius. We have fixed it to 8 pixels as a balance between the error of the ground truth measurement and having a precise background measurement in the exact galaxy location. To select empty stamps for the training sample we identify sources by cross-correlating the sky coordinates of a given image location with the sky coordinates of the sources in the COSMOS catalogue \citep{Laigle}.  \\

In any deep learning algorithm, the training and test samples should be as similar as possible. In our case, our training sample does not contain target galaxies whereas the test sample does.
We therefore add simulated galaxies at the center of the empty training stamps. The simulated galaxies are constructed with parameters based on PAUS data: the Sersic profile parameters, $r_{\rm 50}$, $I_{\rm 50}$ and the magnitude in the $i$-band. The Sersic profile describes the 
surface brightness profile ($I$) of a galaxy. The radius $r$ that contains 50\% of the light intensity ($I_{\rm 50}$) is $r_{\rm 50}$.  These simulated galaxies may differ from the real ones. \LC{For this reason, we mask the central 16 x 16 pixels in both the training and test samples. Although the simulated galaxy is now masked, it is still important to include it, as for some profiles the galaxy light extends outside the masked region. Without the simulated galaxy, \textsc{BKGnet} fails on testing bright sources. As the label is estimated in a 8 pixels radius  aperture, 16 x 16 pixels is the minimum area such that the network does not see the pixels used for the estimation of the ground truth background.}\\

We normalise the stamp before feeding the network. There are different ways of doing this. We apply a normalisation stamp by stamp, where we use the mean and the standard deviation of each stamp to normalise it. We have chosen this normalisation method as it performs better on our dataset. \\
 
We use all the PAUCam images in the COSMOS field to train and validate the network. We have 4928 PAUCam images before the intervention and 4821 after (see sec. \ref{sec:data} for details). For each of them, we sample around 40 stamps per CCD image, giving a total of around 400,000 stamps. We use 90\% of them for training and the remaining 10\% for validation. 

\subsection{Training process and loss function}
Supervised deep learning algorithms are trained comparing the true value with the algorithm's prediction. The agreement between the prediction and the true value is evaluated with a loss function. The choice of loss function depends on the kind of problem one is facing, (e.g. classification, regression). A typical loss function for classification problems is the cross-entropy loss, whereas in regression problems the mean squared error is commonly used. With BKGnet we want the network to associate an uncertainty to each prediction. In supervised deep learning, there are some methods based on Bayesian statistics that deal with uncertainties associated with the predictions  \citep[e.g.][]{ BayesianNN1, BayesianNN2}.\\

The method we use assumes that the distribution $p(\textbf{y}|f^\textbf{w}(\textbf{x}))$ is Gaussian, where $\textbf{y}$ are the background label values, \textbf{x} are the inputs and $f^\textbf{w}(\textbf{x})$ are the network background predictions. Therefore, the loss function is defined 
\begin{equation}
  Loss = \: - \log{p(f^\textbf{w}(\textbf{x}))} \: = \: \frac{(f^\textbf{w}(\textbf{x}) - y)^2}{\sigma^2} + 2\log\sigma.
   \label{loss_function}
\end{equation}
 In this way, we train the network to provide both, the background prediction $f^\textbf{w}(\textbf{x})$ and its associated error $\sigma$.  Notice that the second term on the right hand side prevents the network from predicting a large error that minimizes the first term.\\

With the loss function in Equation \ref{loss_function}, the network provides an error on the quantity $f^\textbf{w}(\textbf{x}) - y$, which has an associated uncertainty $\sigma^2_{\rm pred} +  \sigma^2_{\rm label}$. 
The error on the prediction is therefore
\begin{equation}
    \sigma_{\rm pred} = \sqrt{\sigma^2_{\rm bkgnet} - \sigma^2_{\rm label}}\:,
    \label{pred_error_eq}
\end{equation} where $\sigma^2_{\rm bkgnet}$ is the error provided by the network and $\sigma^2_{\rm label}$ is the error of the background label. The error of the background label is defined in Equation \eqref{laberr}.\\

\LC{\textsc{BKGnet} is trained in 60 epochs with a batch size of 100 stamps using the ADAM optimiser \citep{ADAM} and a learning rate of $10^{\rm -5}$ (see Appendix \ref{A1} for terminology). The training takes about 2 hours using an NVIDIA TITAN V GPU.  }


\section{Testing \textsc{BKGnet} on simulations}

We test the performance of \textsc{BKGnet} with simulated data. We study how well we can predict the background with the network and explore what data are needed and how these data need to be treated before feeding the network. Throughout the rest of the paper we compare the\textsc{BKGnet} predictions to those obtained by calculating the  background inside an annulus around the target source before and after correcting the image with the scattered-light templates.

\subsection{Simulated PAUCam background images}

\LC{The final simulated image $I_{\rm sim}(x,y)$ can be expressed as
\begin{equation}
{\rm I_{\rm sim}}(x,y) = A \cdot \frac{ t_{\rm exp} \cdot {\rm SLT}(x,y) + P(t_{\rm exp} \cdot {\rm SLT}(x,y) )}{t_{\rm exp}},
\label{eq3:sims} 
\end{equation}
where SLT is the scattered light template used to generate the image. In this way, the simulated image shows the same scattered light pattern as the PAUCam images. To generate the Poisson noise, we first multiply with the exposure time $(t_{\rm exp})$ to convert the template from e/s to electrons. Additionally, the template is scaled with a factor $A$ to simulate a wide range of background levels. Finally, we generate a realization of Poisson sky noise $P(\cdot)$ that we add to the image and convert it back to e/s.}\\

\begin{figure}
\includegraphics[width=0.5\textwidth]{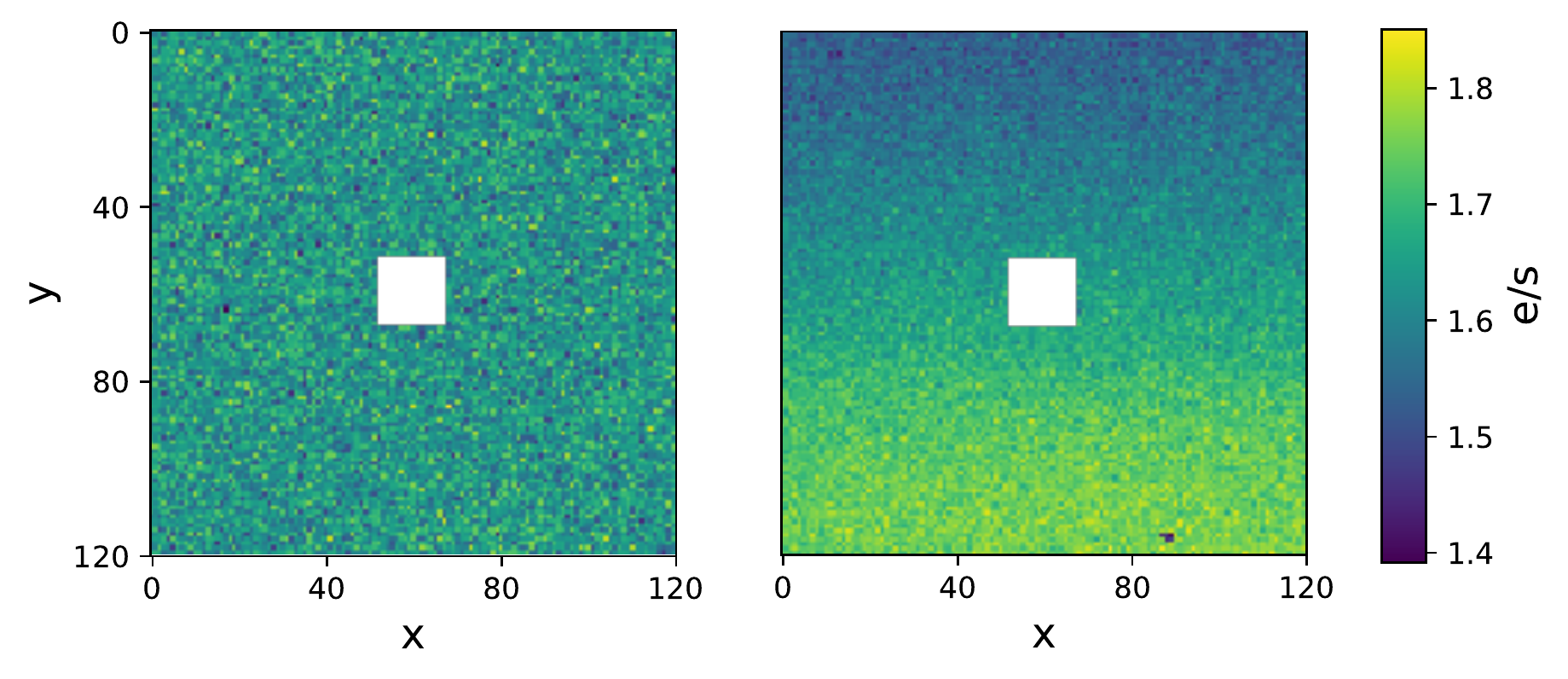}
\centering
\caption{Simulated stamps. On the left, a stamp with a flat background. On the right a stamp with a background with a gradient.}
\label{sim_stamps}
\end{figure}

Figure \ref{sim_stamps} illustrates the effects of scattered light in our simulated stamps. On the left, we show a simulated stamp image with a flat background pattern; and, on the right, with a gradient on the background. This gradient is caused by scattered light (see Fig. \ref{Fig1:scatteredLight_pattern} and \ref{fig3:corrected_image}). Both stamps show a central 8x8 pixel masked region, blocking the light of the galaxy.

\subsection{\textsc{BKGnet} predictions on simulations}

Throughout this section, we train and test on stamps without target galaxies (empty positions). This allows us to test whether it is possible to predict the background with this network's assembly. We also fix the band we are testing to the NB685 filter after the camera intervention. This choice is a compromise between having a considerable amount of scattered light without being completely dominated by it. Before the intervention, the amount of scattered light in some of the CCD images is very large and might not be an adequate choice to test the network. On the other hand, after the intervention, some of the CCDs barely contain scattered light, and those would not be a good choice either. The NB685 filter contains a considerable amount of scattered light and therefore it is a representative example.  We do not need to simulate all bands, as here we only want to test the viability of the the scattered light prediction with \textsc{BKGnet} and to have a better understanding of the network's behaviour. To quantify the background prediction accuracy, we use 
\begin{equation}
    \sigma_{\rm 68} \equiv 0.5 \, (b_{\rm quant}^{84.1} - b_{\rm quant}^{15.9}),
    \label{s68}
\end{equation} 
with quantiles set to 84.1 and 15.9 percentage values. This quantity, $\sigma_{\rm 68}$, is equivalent to a $1\sigma$ error for a normal distribution, but it is less affected by outliers.\\

\begin{figure}
\includegraphics[width=0.5\textwidth]{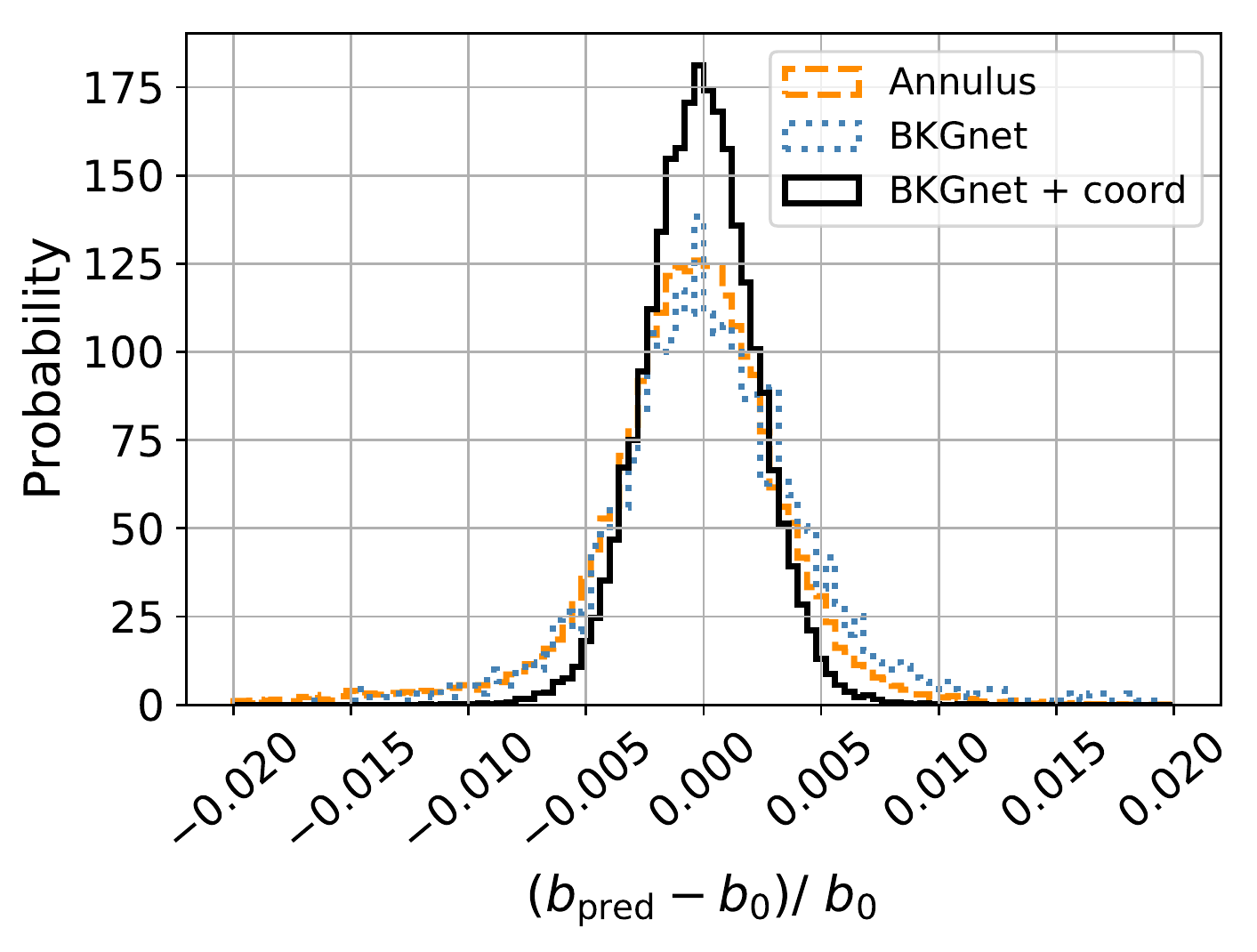}
\centering
\caption{Relative error distributions for the \textsc{BKGnet} (green without coordinate information and orange with coordinate information) and the annulus predictions. $b_{\rm 0}$ is the background label
 and $b_{\rm pred}$ is the background prediction, either for the annulus or for \textsc{BKGnet}.}
\label{fig9:distSims}
\end{figure}

Figure \ref{fig9:distSims} compares the accuracy with which \textsc{BKGnet}
predicts the background to the PAUS default approach. As described before, PAUS estimates the background computing the median inside an annulus centered on the galaxy after the pixel values have been  $\sigma$-clipped.
The plot shows the relative error distribution of the predictions for both methods. We have tested \textsc{BKGnet} with and without embedding the image coordinates of the galaxy. The \textsc{BKGnet} performance improves significantly with the coordinate information. This is not surprising because the amount of scattered light depends on the CCD position
(see Section~\ref{sec:SL_modelling}). Although scattered light is encoded in the image, the CCD position also includes essential information for the prediction. The network might need it to create something similar to the scattered light template. \textsc{BKGnet} achieves a $\sigma_{\rm 68}$ = 0.0038 with information coming only with the stamps. Including the coordinate information, this improves to $\sigma_{\rm 68}$ = 0.0022. Therefore, the network improves by 70\% with the coordinates embedding. The default background estimate shows tails on both sides of the distribution, and yields  $\sigma_{\rm 68}$ = 0.0033, which means \textsc{BKGnet} improves the estimate by 42\%. 
 
\begin{figure*}
    \includegraphics[width=18cm, height=7cm]{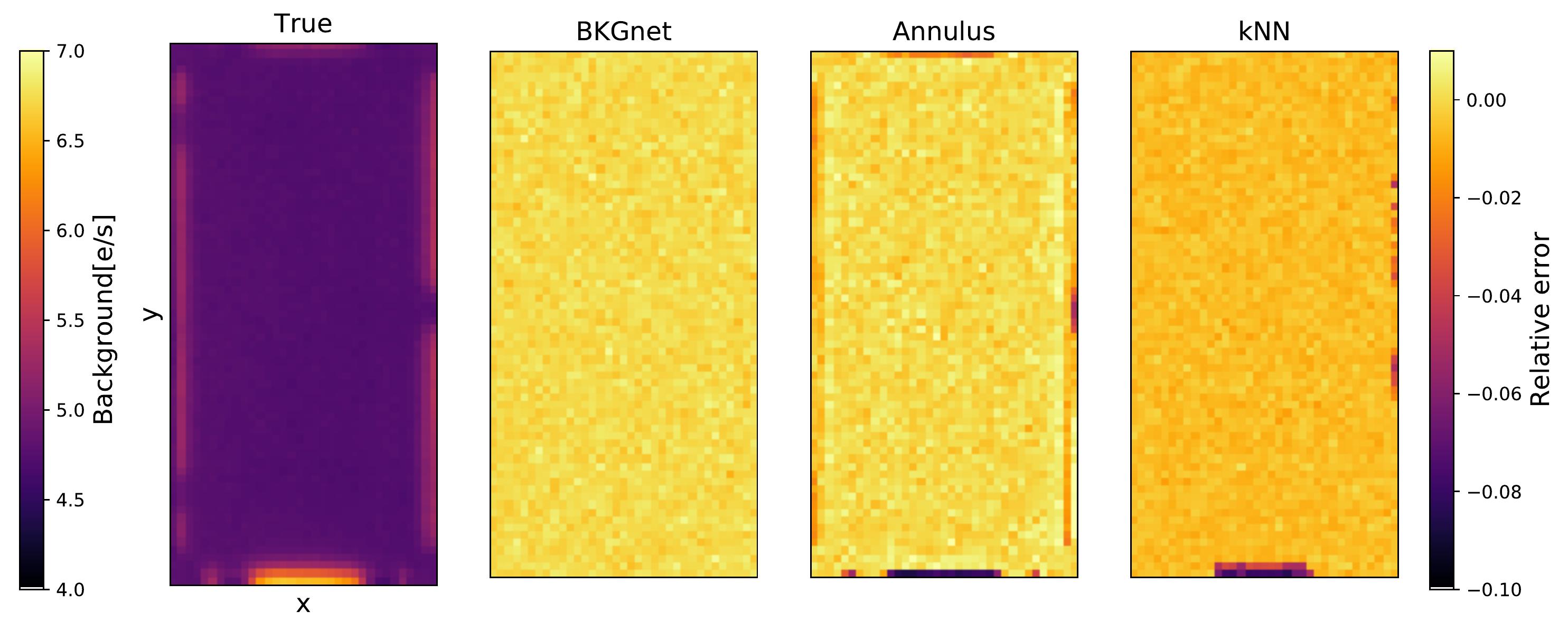}\hfill
    \caption{\emph{Left}: CCD reconstruction with the true background values used to train the network. We sample these background values consecutively and we reconstruct the original image by placing each value in the position it was sampled from. \emph{Second:} Accuracy on the background prediction with \textsc{BKGnet} in the different image positions. We can see there are not spatial patterns. \emph{Third}: Accuracy on the background prediction with the annulus in the different image positions. We can see there are not spatial patterns. \emph{Right}: Accuracy on the background prediction with a kNN in the different image positions.}

    \label{fig10:simsHeatmaps}
\end{figure*}

Figure \ref{fig10:simsHeatmaps} shows the spatial background map (left) and the relative error on the prediction of this map with the annulus background predictions (third panel) and the \textsc{BKGnet} background predictions (second panel). The precision is lower at the edges of the CCD for the annulus-based method, where scattered light is present. This indicates that the tails in Figure~\ref{fig9:distSims} are caused by scattered light. On the other hand, one can see that \textsc{BKGnet} is able to account for the presence of scattered light.\\

\LC{The right panel of Fig.~\ref{fig10:simsHeatmaps} shows the background reconstruction using a kNN algorithm. In addition to the CNN and annulus methods, we have also tested the k-Nearest Neighbors (kNN) \citep{kNN}, Support Vector Regression (SVR) \citep{SVR}, Random Forest (RF) \citep{RF} and a Neural Network (NN) techniques. We used the \textsc{scikit-learn} implementations \citep{scikit-learn} to run the kNN, RF and SVR algorithms. For these tests, unlike the CNN, which can handle images, we input the embedded information and the median pixel value. As shown in Figure \ref{fig10:simsHeatmaps}, the kNN measures the background with less accuracy than \textsc{BKGnet} and the annulus method in the flat background regions. The background measurements are biased by about 3\% , which is 6 times larger than the relative errors (0.5\%). The prediction also shows patterns on the edges with an error 6 times higher than those on flat regions. In contrast, for \textsc{BKGnet} the precision only degrades by a factor of 1.2 when we compare the center with the border positions. Concerning the other methods, the NN provides better predictions than the kNN, although it increases $\sigma_{68}$ by a factor of 2.5 with respect to \textsc{BKGnet}. It also shows patterns on the edges with 4 times higher errors than in flat background regions. Further, the errors of the RF and SVR algorithms are a factor of 6 and 4, respectively, higher than those of the \textsc{BKGnet} method, rendering these methods too imprecise.}  \\

\LC{In the following sections we test the background estimation method on real PAUS images. Based on the performance on simulations, we will only consider the \textsc{BKGnet} and the annulus methods.}


\section{\textsc{BKGnet} on PAUCam images}
\label{sec:resultsPAUCAM}

In the previous section, we have shown that \textsc{BKGnet} is able to accurately predict strongly scattered light backgrounds on simulated blank images including only the background. However,
in real PAUCam data other complications, such as  as cosmic rays, electronic cross-talk, read-out noise and dark current may affect the performance on the estimation of the background. Moreover, correlations between pixels might be introduced during the data reduction process. To examine the impact of these real-life effects we use actual PAUCam images. To assess the accuracy of our measurements, we test network on empty stamps, i.e. without target galaxies.\\

We will use all the images available in COSMOS, splitting the data into those obtained before the camera intervention (in 2016A) and after. This split yields 4928 images and 4821 images for before and after the intervention, respectively. As these numbers are similar, we can easily balance the number of stamps before and after the intervention in our training sample. Although the training sample does not contain target galaxies in the center, sources might be placed in other stamp's positions. To avoid outliers in the training set, e.g. a stamp with a bright star covering most of the background or a bright object too close to the center, we filter the training stamps based on the maximum pixel value. All stamps with a pixel containing more than 100,000 counts are excluded from the training sample.\\

We also exclude 40 images from each subsample before training the network. These 80 images are not used to train the network, but are kept to test it. This is important, as we need to test the network on images it has never seen before. To generate the test set, instead of sampling randomly from the CCDs, we sample stamps consecutively in intervals of 60 pixels. This ensures that we test all CCD regions, including regions affected by scattered light.\\

\begin{figure*}
\includegraphics[width=1\textwidth]{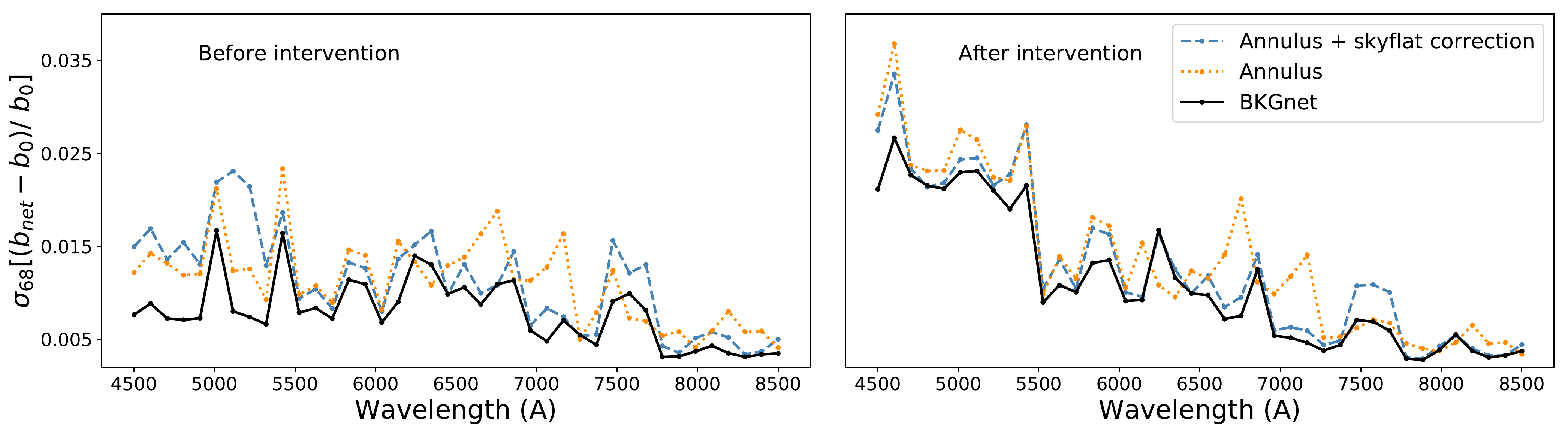}
\centering
\caption{$\sigma_{\rm 68}$ of the relative error in the background prediction for the 40 NBs. \emph{Left}: Before the intervention. \emph{Right}: After the camera intervention. In almost all cases \textsc{BKGnet} performs better than the default approach that employs an annulus to estimate the background.}
\label{fig15:mainResults}
\end{figure*}

Figure \ref{fig15:mainResults} shows the results when we use \textsc{BKGnet} to predict the background on PAUCam images in empty regions. We also show results
when the background is estimated using an annulus, and when we first correct the background variations using a scattered light template ('annulus + sky'). Figure~\ref{fig15:mainResults} shows the value of $\sigma_{\rm 68}$ (Equation \eqref{s68}) of the relative error distribution on the prediction for the 40 different bands. Because we are using the relative error, the comparison between the results before and after the intervention is not representative, as the background levels are different. For instance, in the first filter tray (NB455-NB515), the background before the intervention is between 3 and 5 times higher than after. \\

We focus first on the results before the camera intervention (left panel in Figure \ref{fig15:mainResults}). Images before the camera intervention contain more scattered light than those after (see Fig. \ref{Fig2:intervention}). This makes the scattered-light template modelling more unstable than the modeling of images after the intervention. We find that correcting with the scattered-light template does not improve the annulus method result in every band. In the bluest NBs, i.e. those with the highest amount of scattered light, the scattered-light template seems to decrease the accuracy of the background prediction. On the other hand, \textsc{BKGnet} improves the accuracy compared to the other two methods, especially on the bluest filter tray. On average considering all bands, the network reduces the $\sigma_{\rm 68}$ by $37\%$ compared to the scattered-light template and up to 50\% if we only consider the 8 bluest NBs.\\

If we consider the results after the camera intervention (right panel of Figure \ref{fig15:mainResults}), we see that the scattered-light template improves the annulus method prediction in all the bands. This is expected from the top panel in Figure \ref{Fig2:intervention}, which shows that scattered light trends are stable after the camera intervention. Before the intervention the scattered-light template fails in the bluer bands, which no longer happens after the camera intervention. Nevertheless, \textsc{BKGnet} performs even better: on average, after the intervention it achieves an 18\% improvement compared to the scattered-light template correction. \\

\begin{table}
\center
\begin{tabular}{c|c|c|c|c|}

                                                  & \multicolumn{2}{c|}{BEFORE} & \multicolumn{2}{c|}{AFTER} \\ 
\multicolumn{1}{l|}{}                             & filtered     & sources    & filtered    & sources    \\ \hline
\multicolumn{1}{|c|}{\textbf{Annulus}}            & 0.011          & 0.011      & 0.014         & 0.014      \\ 
\multicolumn{1}{|c|}{\textbf{+ SLT}} & 0.011          & 0.011      & 0.011         & 0.013      \\ 
\multicolumn{1}{|c|}{\textbf{BKGnet}}             & 0.008          & 0.008      & 0.011         & 0.011      \\ \hline
\end{tabular}
\caption{Average $\sigma_{\rm 68}$ of the relative error in the background prediction across all the bands for \textsc{BKGnet} trained before and after the camera intervention. We list the results for the data sets without filtering out stamps affected by sources (`sources'), and if we remove these ('filtered').}
\label{tab1:sig68}
\end{table} 

Table \ref{tab1:sig68} lists the average value of $\sigma_{\rm 68}$ of the relative error in the  background prediction for the three methods: annulus, annulus + scattered-light template and \textsc{BKGnet}. When training and testing, we first exclude the stamps with a maximum pixel value above 100,000. By doing this, we avoid stamps with very bright nearby sources that might bias the prediction. To examine the impact of this step, we also list the results when contaminated stamps are included (`sources')
in Table \ref{tab1:sig68}. These results show that the filtering does not make a difference before the intervention, but the performance improves somewhat for the 
correction that uses the scattered-light templates. The small difference suggests that  scattered light is the main source of bias. For the images without bright sources taken after the intervention, \textsc{BKGnet} and the scattered-light template give the same $\sigma_{\rm 68}$. Therefore, it is possible that \textsc{BKGnet} learns the underlying behaviour of scattered light in a similar way as the scattered-light template. However, as the network also sees the stamp, the correction it infers is more flexible than applying a scattered-light template. This indicates that \textsc{BKGnet} is able to learn how to estimate the background in the presence of other artifacts (e.g. sources or cosmic rays). \\

\textsc{BKGnet} also provides an estimate for the uncertainty associated with the background prediction. To test the accuracy of this estimate we use the empty stamps and study the distribution of  $(b_{\rm net}-b_{\rm true})/\sigma$, where $(b_{\rm net}$ and $\sigma$ are the network predictions. If the errors are correct, this distribution should be a Gaussian with zero mean and unit variance.

\begin{figure}
\includegraphics[width=.5\textwidth]{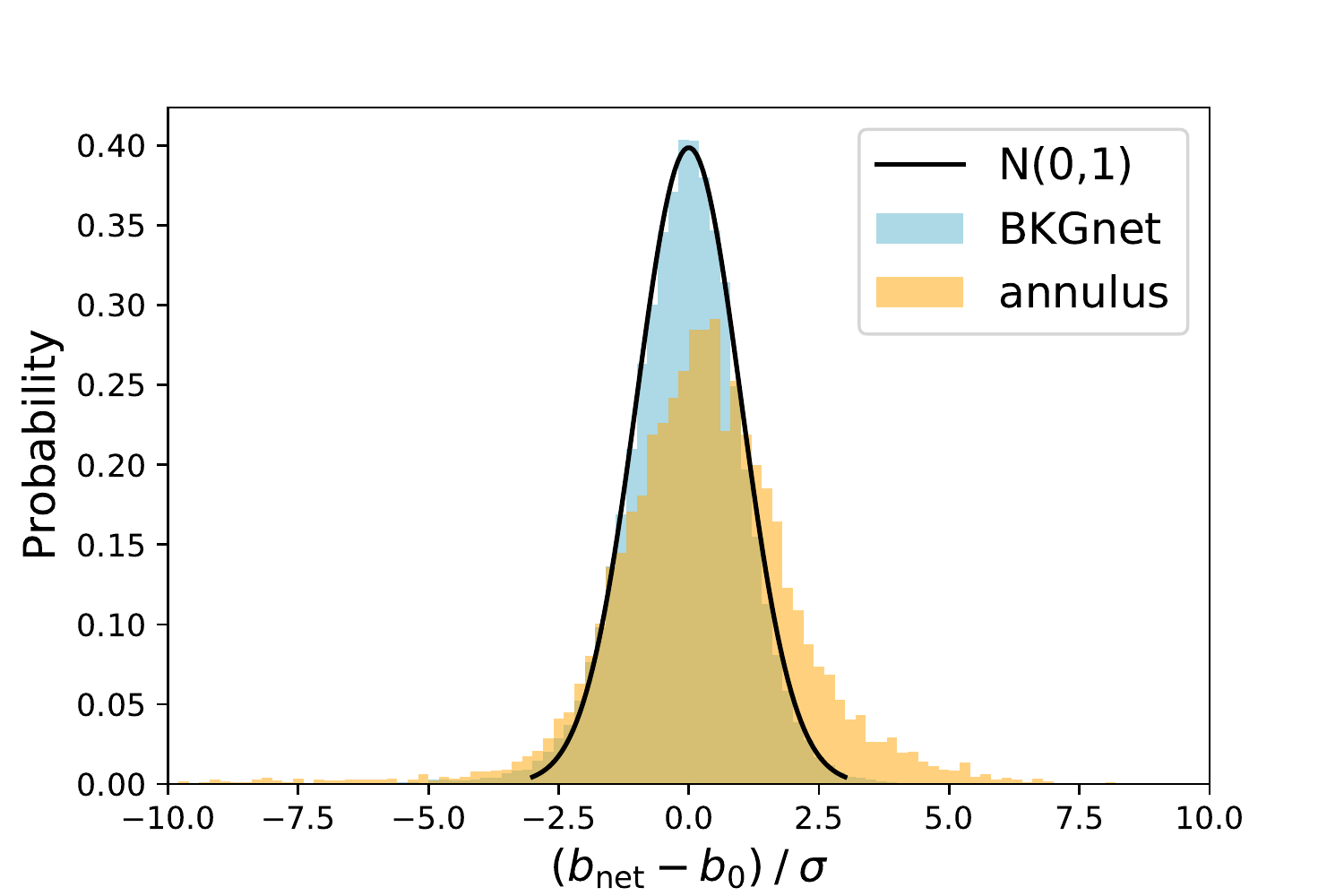}
\centering
\caption{The distribution of $(b_{\rm net} - b_{\rm0})/\sigma$, where $b_{\rm net}$ is the background prediction and $b_{\rm 0}$ the true background. $\sigma$ is the uncertainty in the prediction. We expect the distribution to be a Gaussian centered on zero with unit variance. We show the distribution for the annulus (orange) and \textsc{BKGnet} (blue) predictions.}
\label{fig16:errors}
\end{figure}

Figure \ref{fig16:errors} shows the theoretical Gaussian we should recover and the measured distributions for the annulus and \textsc{BKGnet} predictions. The \textsc{BKGnet} results fit the theoretical Gaussian, which means that our errors are robust. In contrast, the  annulus predictions underestimate the uncertainties by 47\%. Therefore, \textsc{BKGnet} provides a more reliable estimate of the uncertainty in the background determination.

\section{\textsc{BKGnet} validation}

The results presented in section \ref{sec:resultsPAUCAM} show that, compared to the annulus-based methods, \textsc{BKGnet} yields better background estimates (see Fig~\ref{fig15:mainResults} and Tab~\ref{tab1:sig68}). It also provides accurate estimates for the associated uncertainty. However, these tests were done on stamps without galaxies. Here we increase the realism of the problem and quantify the performance of \textsc{BKGnet} at galaxy positions. \\

\subsection{Generating the PAUS catalogue with \textsc{BKGnet} predictions}

We use \textsc{BKGnet} to estimate the background for galaxies in the COSMOS field. We compare the results to those from the \textsc{PAUdm} catalogue, which uses an annulus to determine the background. These catalogues contain around 12 million flux measurements, approximately half of them done on images taken before the intervention and the other half on images taken after the intervention. The galaxy fluxes are obtained subtracting the background from the PAUS raw signal measurements,
\begin{equation}
    F = S - N_{\rm a}b,
    \label{Net_flux}
\end{equation} where $F$ is the net galaxy flux, $S$ is the total signal measured inside the aperture, $N_{\rm a}$ is the number of pixels inside the aperture and $b$ is the predicted background per pixel.  When the background is estimated with an annulus, the error on the net flux is 
\begin{equation}
    \sigma^2 = (S - b) + N_{\rm a} \sigma_{\rm b}^2 + N^2_{\rm a}\Big(\frac{\pi}{2}\Big)\frac{\sigma_{\rm b}^2}{N_{\rm b}}, 
    \label{PAUdm_errorBKG}
\end{equation} where $b$ and $\sigma_{\rm b}$ are the background and the background error in that region and $N_{\rm b}$ is the number of pixels inside the annulus. The $\pi/2$ factor arises from that fact that we use the median of the pixels inside the annulus instead of the mean \footnote{http://wise2.ipac.caltech.edu/staff/fmasci/ApPhotUncert.pdf}.

\noindent For \textsc{BKGnet} the error on the galaxy flux is
\begin{equation}
     \sigma^2 = (S - b) + N_{\rm a} (b + RN^2) + N_{\rm a}^2\sigma_{\rm b}^2 ,
    \label{BKGnet_error}   
\end{equation}where $RN$ is the read-out noise.  \\

Equations \ref{PAUdm_errorBKG} and \ref{BKGnet_error} reflect the differences in the flux uncertainty when the background is measured with an annulus or with \textsc{BKGnet}. In general, there are three main contributors to the flux uncertainty: the uncertainty in the net galaxy flux, the uncertainty in the background estimate, and the uncertainty introduced by the background subtraction. For both background estimation methods we assume that the uncertainty in the net galaxy flux is captured by shot noise. For \textsc{BKGnet}, the background uncertainty is also described by shot noise 
(Eq.~\ref{BKGnet_error}), but we add a read-out noise contribution to the background error. For the \textsc{PAUdm} measurements, the background uncertainty is given by the mean variance per pixel 
(Eq.~\ref{PAUdm_errorBKG}). Therefore, for \textsc{PAUdm}, this term should also account for other error contributions besides shot noise.  The third terms in Eqs.~\ref{PAUdm_errorBKG} and \ref{BKGnet_error} are the contributions from background subtraction uncertainties. In \textsc{PAUdm}, this is determined by the subtraction of a background measured in the annulus. In contrast, in Equation~\ref{BKGnet_error} we use the uncertainty provided by the network within the aperture where the flux is estimated.

\subsection{Validating the catalogues}

The estimation of the background using an annulus is a viable method when the background is flat. In PAUS data, scattered light only affects objects near the edges of the images. Hence, for most of the galaxies in PAUS data the background should be (approximately) flat and we should not expect large differences between the \textsc{BKGnet} and the \textsc{PAUdm} catalogues. Comparing the fluxes estimated  with Equation \ref{Net_flux}, we find a 2\% difference between the two approaches. On the other hand, the uncertainties estimated with \textsc{BKGnet} 
(Eq.~\ref{BKGnet_error}) are 4\% lower than for \textsc{PAUdm}. \\

We need to determine which catalogue provides better photometry estimates. To do so, we use the fact that PAUCam takes multiple observations of the same object in all NB filters. We can compare different exposures of the same object, which should be comparable once the background noise is subtracted. This is formulated as
\begin{equation}
    D \equiv \frac{(e_{\rm 1} - e_{\rm 2})}{\sqrt{(\sigma^2_{\rm 1} + \sigma^2_{\rm 2})}},
    \label{Duplicates}
\end{equation} where $e_i$ are different exposures of the same object and $\sigma_i$ the associated uncertainties. The distribution of $D$ should be a Gaussian with unit variance if the photometry is robust and the errors are properly accounted for. We call this the duplicates test.  \\

\begin{figure}
\includegraphics[width=.5\textwidth]{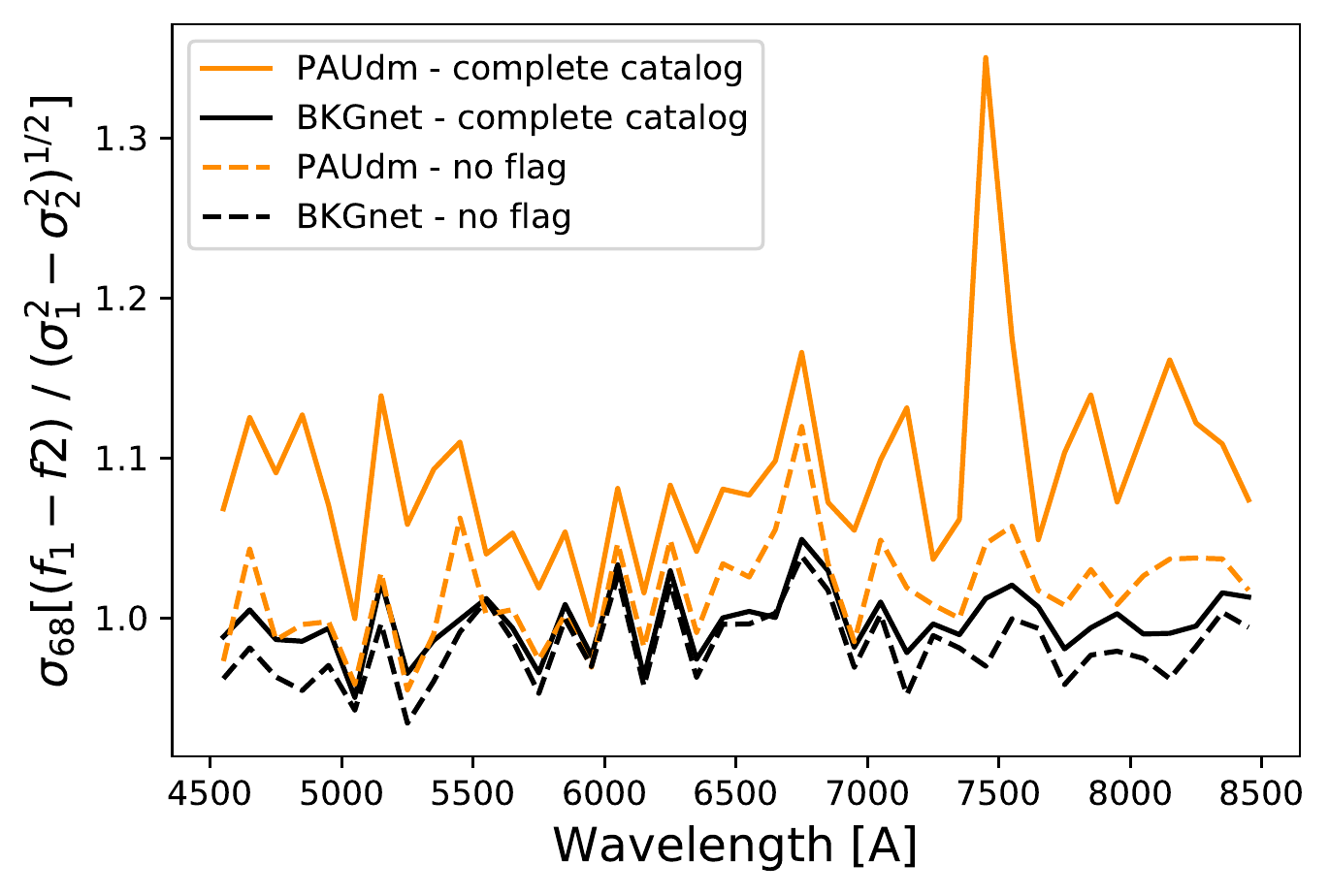}
\centering
\caption{\textsc{BKGnet} validation with the duplicates distribution test. We plot the width of the distribution defined in Equation \eqref{Duplicates} as a function of wavelength for the catalogue generated with \textsc{BKGnet}  (black line) and  the current \textsc{PAUdm} catalogue (orange line). The dashed line corresponds to the results excluding all objects flagged in \textsc{PAUdm}. The solid line includes all objects.}
\label{fig17:Duplicates_mix}
\end{figure}

Figure \ref{fig17:Duplicates_mix} shows the results of the duplicates test as a function of wavelength.  We estimate $\sigma_{\rm 68}[D]$ (Eq. \ref{fig17:Duplicates_mix}) for each NB with the \textsc{BKGnet} (black line) and \textsc{PAUdm} (orange line) catalogues. It is possible to flag photometric outliers based on an ellipticity parameter to detect strongly varying backgrounds. The dashed lines in 
Figure~\ref{fig17:Duplicates_mix} show the results when we exclude such flagged objects. Dropping flagged objects does not significantly change the measurements for \textsc{BKGnet}, but we see a clear improvement for the \textsc{PAUdm} measurements. 
The improvement is particularly prominent for the NB755 filter (at 7500\AA), which is affected by telluric absorption of O$_{\rm 2}$ in the atmosphere.  Interestingly \textsc{BKGnet} seems to know how to deal with these objects, showing that \textsc{BKGnet} is more robust towards various sources of bias, not only scattered light. When we consider all NBs, we find $\langle \sigma_{\rm 68}[D] \rangle = 1.00$ for \textsc{BKGnet}, which is what we would expect for correct photometry. On the other hand, the current \textsc{PAUdm} catalogue yields $\langle\sigma_{\rm 68}[D]\rangle$ = 1.10, i.e. it overestimates the uncertainties.\\

\begin{figure}
\includegraphics[width=.5\textwidth]{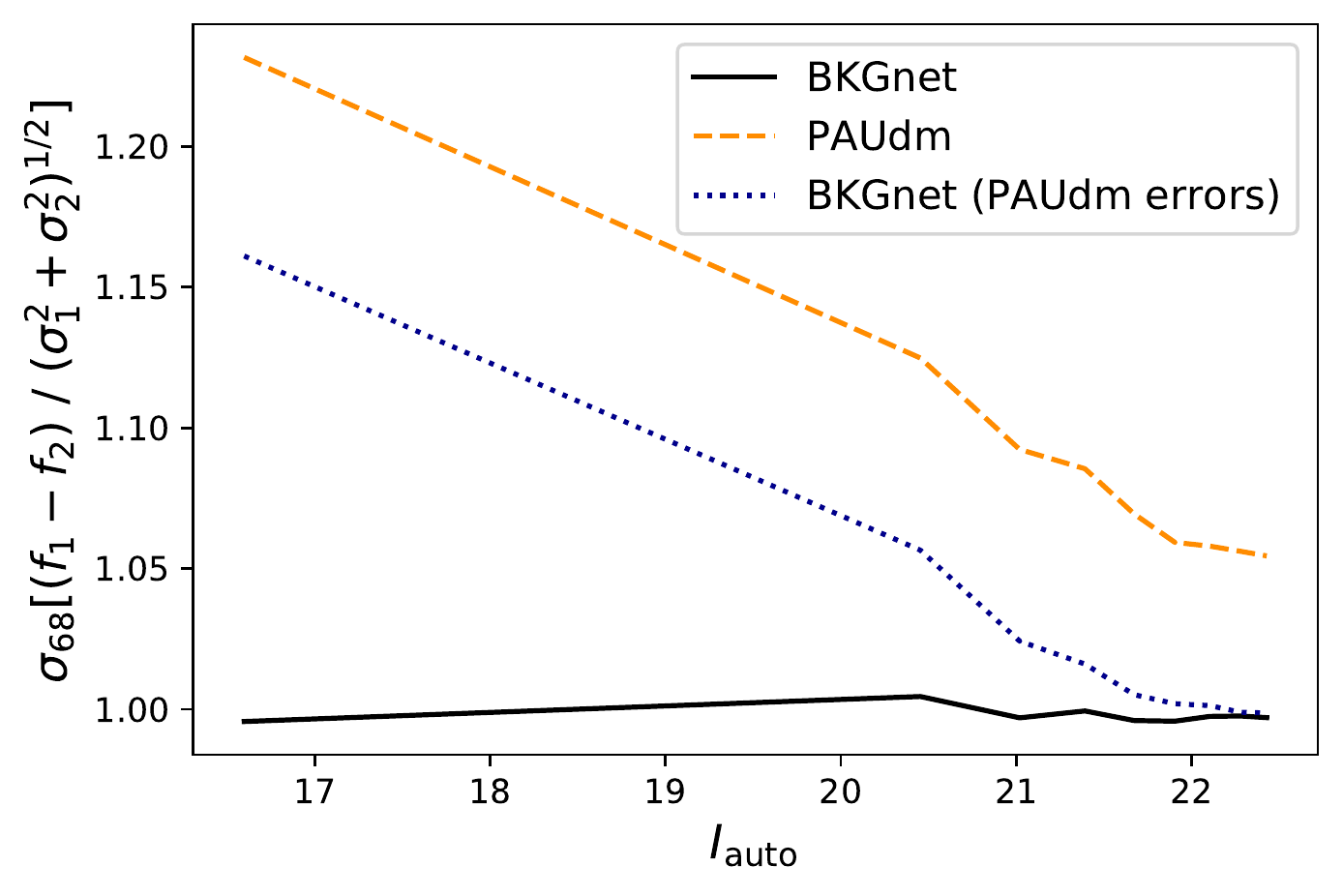}
\centering
\caption{\textsc{BKGnet} validation with the duplicates distribution test. We plot the width of the distribution defined in Equation \eqref{Duplicates} as a function of $i_{\rm auto}$ in the Subaru $i$ band for the catalogue generated with \textsc{BKGnet} (black solid line),  the current \textsc{PAUdm} catalogue (orange dashed line) and a mixed catalogue with the predictions from \textsc{BKGnet} and the errors from \textsc{PAUdm} (blue dotted line).}
\label{fig17:Duplicates_mag}
\end{figure}

The measurement uncertainties should depend on the brightness of the source. To explore this we show $\sigma_{\rm 68}[D]$ as a function of the Subaru $i_{\rm Auto}$ magnitude in Figure~\ref{fig17:Duplicates_mag}.
In the \textsc{PAUdm} catalogue there is a strong trend with magnitude. At the bright end, the fluxes differ by more than 20\% compared to the expectation. This trend disappears  when we predict the background and uncertainties with \textsc{BKGnet}. To explore the origin of the trend further we used the background prediction from \textsc{BKGnet} but the errors from the annulus method. As the blue dotted line in Figure~\ref{fig17:Duplicates_mag} shows, we find the same trend with magnitude. This implies that it is caused by the estimated uncertainties for the annulus method. Moreover, the blue dotted line lies below the \textsc{PAUdm} line. The only difference between these two curves is the background value prediction (not the error). Therefore, the predictions with \textsc{PAUdm} are more accurate than those with the annulus method.\\

To further validate the \textsc{BKGnet} catalogue we run BCNz2 \citep{photoz-Martin} using the fluxes determined using \textsc{BKGnet}. For this test, we exclude the objects flagged in the \textsc{PAUdm} catalogue, in order to use exactly the same objects as in \citep{photoz-Martin}. However, as shown in Figures \ref{fig17:Duplicates_mix} and \ref{fig17:Duplicates_mag}, we do not need to exclude these objects. The photo-$z$s are compared to secure spectroscopic estimates from zCOSMOS DR3 \citep{zCOSMOS} with $i_{\rm AB}<22.5$. We split the sample based on a quality parameter defined as:

\begin{equation}
    {\rm Qz} \equiv \frac{\chi ^2}{N_{\rm f} -3} \Big( \frac{z_{\rm quant}^{\rm 99}- z_{\rm quant}^{\rm 1}}{\rm ODDS(\Delta z = 0.01) }\Big), 
\end{equation} where  $\chi ^2/(n_{\rm f}-1)$ is the reduced chi-squared from the template fit and the $z_{\rm quant}$ are the percentiles of $(z_{\rm photo} -
z_{\rm spec})/(1 + z_{\rm spec}$). The ODDS is defined as 
\begin{equation}
    {\rm ODDS} \equiv \int_{z_{\rm b} - \Delta z}^{z_{\rm b} + \Delta z} {\rm dz\ p(z)}, 
\end{equation} where $z_{\rm b}$ is the mode of the $p(z)$ and $\Delta z$ defines a redshift interval around the peak. In PAUS, a galaxy is considered an outlier if
\begin{equation}
    |z_{\rm  photo} - z_{\rm spec}|\ /\ (1 + z_{\rm spec}) > 0.02.
\end{equation} Notice that this outlier definition is very strict. In broad band photometry, a common outlier definition is $|z_{\rm  photo} - z_{\rm spec}| > 0.15\,(1 + z_{\rm spec})$, e.g. \citet{CFHT_photoz,KiDs_photoz}.\\

Table \ref{tab:photoz} lists the outlier rate and the photometric redshift precision obtained with BCNz2 for the two catalogues. To quantify the redshift precision we use $\sigma_{\rm 68}$ 
(Eq.~\ref{s68}). The photometric redshift precision does not improve significantly between the two catalogues, but we find a reduction in the outlier rate. If we consider the complete sample
(100\%) this improvement is small. This might be because in the full sample the outliers are dominated by photo-$z$ outliers, rather than outliers on the photometry itself. However, if we cut using the $Qz$ parameter to get the best 20\% and 50\% of the sample, we notice that the outlier rate reduces significantly. These outliers should be dominated by photometric outliers. For the best 50\% of the sample we reduce the number of outliers by 25\%, whereas for the best 20\% of objects this improvement rises to a 35\%. This shows once more that \textsc{BKGnet} is a statistically accurate method that is also robust.

\begin{table}
\begin{tabular}{ccccc}
\multicolumn{1}{l}{}           & \multicolumn{2}{c}{\textbf{Outlier percentage}}                       & \multicolumn{2}{c}{\textbf{$10^3\sigma_{\rm 68}$}} \\
\multicolumn{1}{l}{Percentage} & \textsc{BKGnet} & \textsc{PAUdm} & \textsc{BKGnet}                 & \textsc{PAUdm}                 \\\hline
\textbf{20}                             & 3.5                              & 5.4                             & 2.0                                              & 2.1                                             \\ 
\textbf{50}                            & 3.8                              & 5.1                             & 3.6                                              & 3.7                                             \\ 
\textbf{80}                             & 10.4                             & 11.3                            & 5.8                                              & 6.0                                             \\
\textbf{100}                            & 16.7                             & 17.5                            & 8.4                                              & 8.6                                             \\ \hline
\end{tabular}

    \caption{Photo-$z$ outlier rate and accuracy  obtained with BCNz2 for the \textsc{BKGnet} and the \textsc{PAUdm} catalogues. The percentages correspond to the samples selected by the photo-$z$ quality parameters $Qz$.}
    \label{tab:photoz}
\end{table}


\section{Conclusions}

Imaging surveys need accurate background subtraction methods to obtain precise source photometry. We have developed a deep learning method to predict the background for astronomical images that are affected by scattered light. The algorithm has been developed to predict the background on images taken with PAUCam.  The edges of PAUCam images are affected by scattered light (see Fig. \ref{Fig1:scatteredLight_pattern}), especially in the bluer bands. In 2016, the camera was modified to reduce the amount of scattered light. While the amount of scattered light decreased drastically, PAUcam images still contain a significant amount of scattered light (see Fig. \ref{Fig2:intervention}).\\

For each band, the scattered light follows the same spatial pattern within the CCD and scales approximately linear with the background level. We have constructed scattered-light templates and background pixel maps by combining images taken with the same NB and normalised by their background level. These scattered-light templates show the scattered light variation across the CCD and can be used to correct for scattered light (see Fig. \ref{fig3:corrected_image}). Nevertheless, background fluctuations due to external conditions (e.g. Moon, seeing, airmass) can trigger differences on scattered light from night to night. To accurately correct scattered light with scattered-light templates, we would need to generate a scattered-light templates per NB and night. However, even then, fluctuations during the night or a small number of available images in a given band can lead to inaccurate corrections.\\

We therefore developed \textsc{BKGnet}, a deep learning based algorithm that predicts the background and its associated uncertainty behind target sources accounting for scattered light and other distorting effect. \textsc{BKGnet} consists of a  Convolutional Neural Network followed by a linear neural network (see Fig. \ref{fig6:bkgNet_scheme}). In the training set we use empty stamps, i.e. without a target galaxy, so that we can estimate the true  background and use it for training. We need to simulate target galaxies in the training sample before masking the central region, otherwise the network fails when applied to bright and large sources.\\


We first tested the predictions on PAUCam empty stamps, i.e. without target galaxies. For data taken before the intervention, \textsc{BKGnet} improves over the scattered-light templates + annulus prediction by $37\%$. The scattered-light templates correction fails in many of the bands, specially on the bluer filter tray, which is affected the most by scattered light (left panel of Fig.~\ref{fig15:mainResults}). For data taken after the intervention, \textsc{BKGnet} improves over the scattered-light templates + annulus prediction by $17\%$ (right panel of Fig.~\ref{fig15:mainResults}).

\textsc{BKGnet} also predicts the uncertainty associated with the background prediction. For that, we use the log likelihood of a Gaussian centered at the background true value as loss function  (Eq. \ref{loss_function}).  To validate \textsc{BKGnet}, we test on empty positions and estimate the difference between the prediction and the background level, divided by the estimated uncertainty. For the annulus method, we find that the errors are underestimated by 47\% (Fig \ref{fig16:errors}). On the other hand, with \textsc{BKGnet} this quantity is normally distributed around zero with unit variance, showing that the uncertainties are correctly estimated.\\ 

We generated a PAUS catalogue for the COSMOS field using \textsc{BKGnet} to predict the background. To validate the catalogue we took advantage of having multiple measurements of the same object.  The resulting distribution of differences in flux measurements should be a Gaussian of unit variance (Eq. \ref{Duplicates}). The results demonstrate that \textsc{BKGnet} improves the photometry with respect to the current background subtraction algorithm. We test the performance for the full catalogue and a catalogue where we exclude all objects flagged in the current catalogue version. When excluding flagged objects, we find very similar results with the \textsc{BKGnet} catalogue and the current catalogue. However, when testing the full catalogue, we find a large improvement for \textsc{BKGnet}. It specially improves the results in a region with high atmospheric absorption, demonstrating that it is more robust against sources of bias while still being statistically accurate. It also removes a strong systematic trend with $i$-band magnitude, that disappears when the uncertainties are estimated with the network. \\

Finally, as the aim of PAUS is to provide accurate redshifts for large samples of galaxies, we have run the \textsc{BCNz2} code using the \textsc{BKGnet} catalogue. \textsc{BKGnet} reduces the outlier rate by a 25\% and 35\% respectively for the best 50\% and 20\% photo-$z$ samples, while the accuracy is not affected.\\

\LC{With \textsc{BKGnet} we have optimised the background subtraction task, one of the image processing steps in photometric surveys that can improve the redshift estimation and classification of galaxies. Deep learning algorithms that predict these quantities directly from images  have to subtract the background internally. Therefore, the understanding from \textsc{BKGnet} will also help to optimise such deep learning algorithms. Although the network has been tested with PAUCam images, the concept should also be applicable to future imaging surveys as {\it Euclid} and LSST.}

\section*{Acknowledgement}
 Funding for PAUS has been provided by Durham University (via the ERC StG DEGAS-259586), ETH Zurich, Leiden University (via ERC StG ADULT-279396 and Netherlands Organisation for Scientific Research (NWO) Vici grant 639.043.512) and University College London. The PAUS participants from Spanish institutions are partially supported by MINECO under grants CSD2007-00060, AYA2015-71825, ESP2015-88861, FPA2015-68048, SEV-2016-0588, SEV-2016-0597, and MDM-2015-0509, some of which include ERDF funds from the European Union. IEEC and IFAE are partially funded by the CERCA program of the Generalitat de Catalunya. The PAU data center is hosted by the Port d'Informaci\'o Cient\'ifica (PIC), maintained through a collaboration of CIEMAT and IFAE, with additional support from Universitat Aut\`onoma de Barcelona and ERDF. CosmoHub has been developed by PIC and was partially funded by the "Plan Estatal de Investigaci\'on Cient\'ifica y T\'ecnica y de Innovaci\'on" program of the Spanish government. We gratefully acknowledge the support of NVIDIA Corporation with the donation of the Titan V  GPU used for this research.
 This project has received funding from the European Union's Horizon 2020 research and innovation programme under grant agreement No 776247. Adam Amara is supported by a Royal Society Wolfson Fellowship. MS has been supported by the National Science Centre (grant UMO-2016/23/N/ST9/02963).
\bibliographystyle{mnras}
\bibliography{BKG}

\begin{thebibliography}{}
\makeatletter
\relax
\def\mn@urlcharsother{\let\do\@makeother \do\$\do\&\do\#\do\^\do\_\do\%\do\~}
\def\mn@doi{\begingroup\mn@urlcharsother \@ifnextchar [ {\mn@doi@}
  {\mn@doi@[]}}
\def\mn@doi@[#1]#2{\def\@tempa{#1}\ifx\@tempa\@empty \href
  {http://dx.doi.org/#2} {doi:#2}\else \href {http://dx.doi.org/#2} {#1}\fi
  \endgroup}
\def\mn@eprint#1#2{\mn@eprint@#1:#2::\@nil}
\def\mn@eprint@arXiv#1{\href {http://arxiv.org/abs/#1} {{\tt arXiv:#1}}}
\def\mn@eprint@dblp#1{\href {http://dblp.uni-trier.de/rec/bibtex/#1.xml}
  {dblp:#1}}
\def\mn@eprint@#1:#2:#3:#4\@nil{\def\@tempa {#1}\def\@tempb {#2}\def\@tempc
  {#3}\ifx \@tempc \@empty \let \@tempc \@tempb \let \@tempb \@tempa \fi \ifx
  \@tempb \@empty \def\@tempb {arXiv}\fi \@ifundefined
  {mn@eprint@\@tempb}{\@tempb:\@tempc}{\expandafter \expandafter \csname
  mn@eprint@\@tempb\endcsname \expandafter{\@tempc}}}

\bibitem[\protect\citeauthoryear{{Abbott} et~al.,}{{Abbott}
  et~al.}{2018}]{DESDR1}
{Abbott} T.~M.~C.,  et~al., 2018, \mn@doi [\apjs] {10.3847/1538-4365/aae9f0},
  \href {https://ui.adsabs.harvard.edu/abs/2018ApJS..239...18A} {239, 18}

\bibitem[\protect\citeauthoryear{{Alexander}, {Gleyzer}, {McDonough}, {Toomey}
  \& {Usai}}{{Alexander} et~al.}{2019}]{class}
{Alexander} S.,  {Gleyzer} S.,  {McDonough} E.,  {Toomey} M.~W.,   {Usai} E.,
  2019, \href {https://ui.adsabs.harvard.edu/abs/2019arXiv190907346A} {p.
  arXiv:1909.07346}

\bibitem[\protect\citeauthoryear{{Bertin} \& {Arnouts}}{{Bertin} \&
  {Arnouts}}{1996}]{SExtractor}
{Bertin} E.,  {Arnouts} S.,  1996, \mn@doi [Astronomy and Astrophysics
  Supplement Series] {10.1051/aas:1996164}, \href
  {https://ui.adsabs.harvard.edu/\#abs/1996A&AS..117..393B} {117, 393}

\bibitem[\protect\citeauthoryear{{Bijaoui}}{{Bijaoui}}{1980}]{Bijaoui}
{Bijaoui} A.,  1980, \aap, \href
  {https://ui.adsabs.harvard.edu/\#abs/1980A&A....84...81B} {84, 81}

\bibitem[\protect\citeauthoryear{{Bilicki} et~al.,}{{Bilicki}
  et~al.}{2018}]{KiDs_photoz}
{Bilicki} M.,  et~al., 2018, \mn@doi [\aap] {10.1051/0004-6361/201731942},
  \href {https://ui.adsabs.harvard.edu/abs/2018A&A...616A..69B} {616, A69}

\bibitem[\protect\citeauthoryear{Breiman}{Breiman}{2001}]{RF}
Breiman L.,  2001, \mn@doi [Mach. Learn.] {10.1023/A:1010933404324}, 45, 5

\bibitem[\protect\citeauthoryear{{Cabayol} et~al.,}{{Cabayol}
  et~al.}{2019}]{Cabayol}
{Cabayol} L.,  et~al., 2019, \mn@doi [\mnras] {10.1093/mnras/sty3129}, \href
  {https://ui.adsabs.harvard.edu/\#abs/2019MNRAS.483..529C} {483, 529}

\bibitem[\protect\citeauthoryear{{Carrasco-Davis} et~al.,}{{Carrasco-Davis}
  et~al.}{2018}]{objseq_class}
{Carrasco-Davis} R.,  et~al., 2018, \href
  {https://ui.adsabs.harvard.edu/abs/2018arXiv180703869C} {p. arXiv:1807.03869}

\bibitem[\protect\citeauthoryear{{Casas} et~al.,}{{Casas}
  et~al.}{2012}]{Casas2012}
{Casas} R.,  et~al., 2012, in High Energy, Optical, and Infrared Detectors for
  Astronomy V. p. 845326, \mn@doi{10.1117/12.924640}

\bibitem[\protect\citeauthoryear{{Casas} et~al.,}{{Casas}
  et~al.}{2016}]{Casas2016}
{Casas} R.,  et~al., 2016, in Ground-based and Airborne Instrumentation for
  Astronomy VI. p. 99084K, \mn@doi{10.1117/12.2232422}

\bibitem[\protect\citeauthoryear{{Castander} et~al.,}{{Castander}
  et~al.}{2012}]{PAUCAM_Francisco}
{Castander} F.~J.,  et~al., 2012, in Ground-based and Airborne Instrumentation
  for Astronomy IV. p. 84466D, \mn@doi{10.1117/12.926234}

\bibitem[\protect\citeauthoryear{{Castander}, {Eriksen}, {Serrano}  \& {et
  al.}}{{Castander} et~al.}{prep}]{PAUcalib}
{Castander} F.,  {Eriksen} M.,  {Serrano} S.,   {et al.} {in prep.}

\bibitem[\protect\citeauthoryear{Cover \& Hart}{Cover \& Hart}{2006}]{kNN}
Cover T.,  Hart P.,  2006, \mn@doi [IEEE Trans. Inf. Theor.]
  {10.1109/TIT.1967.1053964}, 13, 21

\bibitem[\protect\citeauthoryear{{D'Isanto} \& {Polsterer}}{{D'Isanto} \&
  {Polsterer}}{2018}]{photoz}
{D'Isanto} A.,  {Polsterer} K.~L.,  2018, \mn@doi [\aap]
  {10.1051/0004-6361/201731326}, \href
  {https://ui.adsabs.harvard.edu/abs/2018A&A...609A.111D} {609, A111}

\bibitem[\protect\citeauthoryear{Drucker, Burges, Kaufman, Smola  \&
  Vapnik}{Drucker et~al.}{1996}]{SVR}
Drucker H.,  Burges C. J.~C.,  Kaufman L.,  Smola A.,   Vapnik V.,  1996,
  Support Vector Regression Machines

\bibitem[\protect\citeauthoryear{{Eriksen} et~al.,}{{Eriksen}
  et~al.}{2019}]{photoz-Martin}
{Eriksen} M.,  et~al., 2019, \mn@doi [\mnras] {10.1093/mnras/stz204}, \href
  {https://ui.adsabs.harvard.edu/abs/2019MNRAS.484.4200E} {484, 4200}

\bibitem[\protect\citeauthoryear{{Fluri}, {Kacprzak}, {Refregier}, {Amara},
  {Lucchi}  \& {Hofmann}}{{Fluri} et~al.}{2018}]{DL_mapsConstrains}
{Fluri} J.,  {Kacprzak} T.,  {Refregier} A.,  {Amara} A.,  {Lucchi} A.,
  {Hofmann} T.,  2018, \mn@doi [\prd] {10.1103/PhysRevD.98.123518}, \href
  {https://ui.adsabs.harvard.edu/\#abs/2018PhRvD..98l3518F} {98, 123518}

\bibitem[\protect\citeauthoryear{{Gaia Collaboration} et~al.,}{{Gaia
  Collaboration} et~al.}{2018}]{GaiaDR2}
{Gaia Collaboration} et~al., 2018, \mn@doi [\aap]
  {10.1051/0004-6361/201833051}, \href
  {https://ui.adsabs.harvard.edu/abs/2018A&A...616A...1G} {616, A1}

\bibitem[\protect\citeauthoryear{{George} \& {Huerta}}{{George} \&
  {Huerta}}{2018}]{GW}
{George} D.,  {Huerta} E.~A.,  2018, \mn@doi [Physics Letters B]
  {10.1016/j.physletb.2017.12.053}, \href
  {https://ui.adsabs.harvard.edu/\#abs/2018PhLB..778...64G} {778, 64}

\bibitem[\protect\citeauthoryear{{Herbel}, {Kacprzak}, {Amara}, {Refregier}  \&
  {Lucchi}}{{Herbel} et~al.}{2018a}]{ETH2}
{Herbel} J.,  {Kacprzak} T.,  {Amara} A.,  {Refregier} A.,   {Lucchi} A.,
  2018a, \mn@doi [\jcap] {10.1088/1475-7516/2018/07/054}, \href
  {https://ui.adsabs.harvard.edu/abs/2018JCAP...07..054H} {2018, 054}

\bibitem[\protect\citeauthoryear{{Herbel}, {Kacprzak}, {Amara}, {Refregier}  \&
  {Lucchi}}{{Herbel} et~al.}{2018b}]{psf}
{Herbel} J.,  {Kacprzak} T.,  {Amara} A.,  {Refregier} A.,   {Lucchi} A.,
  2018b, \mn@doi [\jcap] {10.1088/1475-7516/2018/07/054}, \href
  {https://ui.adsabs.harvard.edu/abs/2018JCAP...07..054H} {2018, 054}

\bibitem[\protect\citeauthoryear{{Ilbert} et~al.,}{{Ilbert}
  et~al.}{2006}]{CFHT_photoz}
{Ilbert} O.,  et~al., 2006, \mn@doi [\aap] {10.1051/0004-6361:20065138}, \href
  {https://ui.adsabs.harvard.edu/abs/2006A&A...457..841I} {457, 841}

\bibitem[\protect\citeauthoryear{{Ivezi{\'c}} et~al.,}{{Ivezi{\'c}}
  et~al.}{2019}]{LSST}
{Ivezi{\'c}} {\v{Z}}.,  et~al., 2019, \mn@doi [\apj]
  {10.3847/1538-4357/ab042c}, \href
  {https://ui.adsabs.harvard.edu/abs/2019ApJ...873..111I} {873, 111}

\bibitem[\protect\citeauthoryear{{Kendall} \& {Gal}}{{Kendall} \&
  {Gal}}{2017}]{BayesianNN1}
{Kendall} A.,  {Gal} Y.,  2017, \href
  {https://ui.adsabs.harvard.edu/abs/2017arXiv170304977K} {p. arXiv:1703.04977}

\bibitem[\protect\citeauthoryear{{Kendall}, {Gal}  \& {Cipolla}}{{Kendall}
  et~al.}{2017}]{BayesianNN2}
{Kendall} A.,  {Gal} Y.,   {Cipolla} R.,  2017, \href
  {https://ui.adsabs.harvard.edu/abs/2017arXiv170507115K} {p. arXiv:1705.07115}

\bibitem[\protect\citeauthoryear{{Kingma} \& {Ba}}{{Kingma} \&
  {Ba}}{2014}]{ADAM}
{Kingma} D.~P.,  {Ba} J.,  2014, arXiv e-prints, \href
  {https://ui.adsabs.harvard.edu/abs/2014arXiv1412.6980K} {p. arXiv:1412.6980}

\bibitem[\protect\citeauthoryear{Krizhevsky, Sutskever  \& Hinton}{Krizhevsky
  et~al.}{2012}]{Alexnet}
Krizhevsky A.,  Sutskever I.,   Hinton G.~E.,  2012, in Proceedings of the 25th
  International Conference on Neural Information Processing Systems - Volume 1.
  NIPS'12.
Curran Associates Inc., USA, pp 1097--1105, \url
  {http://dl.acm.org/citation.cfm?id=2999134.2999257}

\bibitem[\protect\citeauthoryear{{Laigle} et~al.,}{{Laigle}
  et~al.}{2016}]{Laigle}
{Laigle} C.,  et~al., 2016, \mn@doi [\apjs] {10.3847/0067-0049/224/2/24}, \href
  {https://ui.adsabs.harvard.edu/abs/2016ApJS..224...24L} {224, 24}

\bibitem[\protect\citeauthoryear{{Laureijs} et~al.,}{{Laureijs}
  et~al.}{2011}]{Euclid}
{Laureijs} R.,  et~al., 2011, \href
  {https://ui.adsabs.harvard.edu/\#abs/2011arXiv1110.3193L} {p.
  arXiv:1110.3193}

\bibitem[\protect\citeauthoryear{LeCun, Boser, Denker, Henderson, Howard,
  Hubbard  \& Jackel}{LeCun et~al.}{1989}]{lecun89}
LeCun Y.,  Boser B.,  Denker J.~S.,  Henderson D.,  Howard R.~E.,  Hubbard W.,
   Jackel L.~D.,  1989, \mn@doi [Neural Computation]
  {10.1162/neco.1989.1.4.541}, 1, 541

\bibitem[\protect\citeauthoryear{Lecun, Bottou, Bengio  \& Haffner}{Lecun
  et~al.}{1998}]{lecun98}
Lecun Y.,  Bottou L.,  Bengio Y.,   Haffner P.,  1998, \mn@doi [Proceedings of
  the IEEE] {10.1109/5.726791}, 86, 2278

\bibitem[\protect\citeauthoryear{{Lilly} et~al.,}{{Lilly}
  et~al.}{2007}]{zCOSMOS}
{Lilly} S.~J.,  et~al., 2007, \mn@doi [\apjs] {10.1086/516589}, \href
  {https://ui.adsabs.harvard.edu/abs/2007ApJS..172...70L} {172, 70}

\bibitem[\protect\citeauthoryear{{Mart{\'\i}}, {Miquel}, {Castander},
  {Gazta{\~n}aga}, {Eriksen}  \& {S{\'a}nchez}}{{Mart{\'\i}}
  et~al.}{2014}]{photoz-Marti}
{Mart{\'\i}} P.,  {Miquel} R.,  {Castander} F.~J.,  {Gazta{\~n}aga} E.,
  {Eriksen} M.,   {S{\'a}nchez} C.,  2014, \mn@doi [\mnras]
  {10.1093/mnras/stu801}, \href
  {https://ui.adsabs.harvard.edu/\#abs/2014MNRAS.442...92M} {442, 92}

\bibitem[\protect\citeauthoryear{{Newell}}{{Newell}}{1983}]{Newell}
{Newell} E.~B.,  1983, in Astronomical Measuring Machines Workshop. pp 15--40

\bibitem[\protect\citeauthoryear{{Padilla} et~al.,}{{Padilla}
  et~al.}{2016}]{PAUCam-Padilla}
{Padilla} C.,  et~al., 2016, in Ground-based and Airborne Instrumentation for
  Astronomy VI. p. 99080Z, \mn@doi{10.1117/12.2231884}

\bibitem[\protect\citeauthoryear{{Padilla} et~al.,}{{Padilla}
  et~al.}{2019}]{PAUCam}
{Padilla} C.,  et~al., 2019, \mn@doi [\aj] {10.3847/1538-3881/ab0412}, \href
  {https://ui.adsabs.harvard.edu/abs/2019AJ....157..246P} {157, 246}

\bibitem[\protect\citeauthoryear{{Pasquet}, {Bertin}, {Treyer}, {Arnouts}  \&
  {Fouchez}}{{Pasquet} et~al.}{2019}]{photoz_img}
{Pasquet} J.,  {Bertin} E.,  {Treyer} M.,  {Arnouts} S.,   {Fouchez} D.,  2019,
  \mn@doi [\aap] {10.1051/0004-6361/201833617}, \href
  {https://ui.adsabs.harvard.edu/abs/2019A&A...621A..26P} {621, A26}

\bibitem[\protect\citeauthoryear{Paszke et~al.,}{Paszke et~al.}{2017}]{pytorch}
Paszke A.,  et~al., 2017

\bibitem[\protect\citeauthoryear{Pedregosa et~al.,}{Pedregosa
  et~al.}{2011}]{scikit-learn}
Pedregosa F.,  et~al., 2011, Journal of Machine Learning Research, 12, 2825

\bibitem[\protect\citeauthoryear{{Popowicz} \& {Smolka}}{{Popowicz} \&
  {Smolka}}{2015}]{Popowicz}
{Popowicz} A.,  {Smolka} B.,  2015, \mn@doi [\mnras] {10.1093/mnras/stv1320},
  \href {https://ui.adsabs.harvard.edu/\#abs/2015MNRAS.452..809P} {452, 809}

\bibitem[\protect\citeauthoryear{Romanishin}{Romanishin}{2014}]{photometrybook}
Romanishin W.,  2014, An Introduction to Astronomical Photometry Using Ccds.
Createspace Independent Pub, \url
  {https://books.google.es/books?id=0nbMoQEACAAJ}

\bibitem[\protect\citeauthoryear{{Serrano}, {Castander}, {Fernandez}  \& {et
  al.}}{{Serrano} et~al.}{prep}]{PAUimage}
{Serrano} S.,  {Castander} F.,  {Fernandez} E.,   {et al.} in prep.

\bibitem[\protect\citeauthoryear{{Stetson}}{{Stetson}}{1987}]{DAOPHOT}
{Stetson} P.~B.,  1987, \mn@doi [Publications of the Astronomical Society of
  the Pacific] {10.1086/131977}, \href
  {https://ui.adsabs.harvard.edu/\#abs/1987PASP...99..191S} {99, 191}

\bibitem[\protect\citeauthoryear{{Stothert} et~al.,}{{Stothert}
  et~al.}{2018}]{Stothert}
{Stothert} L.,  et~al., 2018, \mn@doi [\mnras] {10.1093/mnras/sty2491}, \href
  {https://ui.adsabs.harvard.edu/\#abs/2018MNRAS.481.4221S} {481, 4221}

\bibitem[\protect\citeauthoryear{Teeninga, Moschini, Trager  \&
  Wilkinson}{Teeninga et~al.}{2015}]{Teeninga}
Teeninga P.,  Moschini U.,  Trager S.~C.,   Wilkinson M. H.~F.,  2015, in 2015
  IEEE International Conference on Image Processing (ICIP). pp 1046--1050,
  \mn@doi{10.1109/ICIP.2015.7350959}

\bibitem[\protect\citeauthoryear{{Tonello} et~al.,}{{Tonello}
  et~al.}{2019}]{Tonello}
{Tonello} N.,  et~al., 2019, \mn@doi [Astronomy and Computing]
  {10.1016/j.ascom.2019.04.002}, \href
  {https://ui.adsabs.harvard.edu/abs/2019A&C....27..171T} {27, 171}

\bibitem[\protect\citeauthoryear{{Tortorelli} et~al.,}{{Tortorelli}
  et~al.}{2018}]{Tortorelli2018}
{Tortorelli} L.,  et~al., 2018, \href
  {http://adsabs.harvard.edu/abs/2018arXiv180505340T} {p. arXiv:1805.05340}

\bibitem[\protect\citeauthoryear{{Vafaei Sadr}, {Vos}, {Bassett}, {Hosenie},
  {Oozeer}  \& {Lochner}}{{Vafaei Sadr} et~al.}{2019}]{DeepSource}
{Vafaei Sadr} A.,  {Vos} E.~E.,  {Bassett} B.~A.,  {Hosenie} Z.,  {Oozeer} N.,
   {Lochner} M.,  2019, \mn@doi [\mnras] {10.1093/mnras/stz131}, \href
  {https://ui.adsabs.harvard.edu/abs/2019MNRAS.484.2793V} {484, 2793}

\bibitem[\protect\citeauthoryear{Voulodimos, Doulamis, Doulamis  \&
  Protopapadakis}{Voulodimos et~al.}{2018}]{computervision}
Voulodimos A.,  Doulamis N.,  Doulamis A.,   Protopapadakis E.,  2018, \mn@doi
  [Computational Intelligence and Neuroscience] {10.1155/2018/7068349}, 2018, 1

\bibitem[\protect\citeauthoryear{Werbos}{Werbos}{1982}]{werbos}
Werbos P.~J.,  1982, in Drenick R.~F.,  Kozin F.,  eds, System Modeling and
  Optimization. Springer Berlin Heidelberg, Berlin, Heidelberg, pp 762--770

\bibitem[\protect\citeauthoryear{{Xu}, {Wang}, {Chen}  \& {Li}}{{Xu}
  et~al.}{2015}]{LReLu}
{Xu} B.,  {Wang} N.,  {Chen} T.,   {Li} M.,  2015, \href
  {https://ui.adsabs.harvard.edu/\#abs/2015arXiv150500853X} {p.
  arXiv:1505.00853}

\bibitem[\protect\citeauthoryear{{Zeiler} \& {Fergus}}{{Zeiler} \&
  {Fergus}}{2013a}]{cnn_understanding}
{Zeiler} M.~D.,  {Fergus} R.,  2013a, \href
  {https://ui.adsabs.harvard.edu/\#abs/2013arXiv1311.2901Z} {p.
  arXiv:1311.2901}

\bibitem[\protect\citeauthoryear{{Zeiler} \& {Fergus}}{{Zeiler} \&
  {Fergus}}{2013b}]{CNN_vis}
{Zeiler} M.~D.,  {Fergus} R.,  2013b, \href
  {https://ui.adsabs.harvard.edu/abs/2013arXiv1311.2901Z} {p. arXiv:1311.2901}

\bibitem[\protect\citeauthoryear{{Zhang} \& {Bloom}}{{Zhang} \&
  {Bloom}}{2019}]{deepCR}
{Zhang} K.,  {Bloom} J.,  2019, \mn@doi [The Journal of Open Source Software]
  {10.21105/joss.01651}, \href
  {https://ui.adsabs.harvard.edu/abs/2019JOSS....4.1651Z} {4, 1651}

\makeatother
\end{thebibliography}
\begin{appendices}
\section{Convolutional Neural Networks}
\label{A1}
Machine learning methods are data analysis techniques where the algorithm learns from the data. In particular, one of the most popular class of algorithms are neural networks \citep{werbos}, which are designed to recognise patterns, usually learned from training data (\textit{supervised method}). They are mainly used for regression and classification problems \citep{class}. Deep learning is a subset of machine learning that refers to a development of neural network technology, involving a large number of layers. \\

Deep learning methods, and in general any supervised machine learning method, model a problem by optimising a set of trainable weights that fit the data. This is done in three stages: \textit{forward propagation}, \textit{back propagation} and \textit{weight optimisation}. The network starts with the \textit{forward propagation}. At this stage, the input data propagates through all the network layers and then, the network gives a prediction for each of the input samples. After that, by comparing with the known true value, which is technically called \textit{label}, the network estimates a prediction error with a given loss function. After that, \textit{back propagation} takes place. \textit{Back propagation} consists of computing the contribution of each weight on the prediction  error. Such contributions are calculated with the partial derivative of the loss with respect to each of the weights. The \textit{weight optimisation} is the weights correction based on the quantities calculated in the \textit{back propagation} to reduce the error in the next iteration.  \\

In this work, we use a Convolutional Neural Network \citep[CNN;][]{lecun98,cnn_understanding}. Our network contains four differentiated types of layers: 
\begin{description}
    \item [\emph{Convolutional layer}:]  This layer makes the network powerful in image and pattern recognition tasks. It has a filter, technically named \textit{kernel} and is usually 2-dimensional, which contains a set of trainable weights used to convolve the image. The outcome of this layer is the input image convolved with the kernel. In a given convolutional layer, one can convolve the input with as many kernels as desired. Each of these convolutions will generate a convolved image, which we refer to as \textit{channel}. All of them together are the input of the next layer.
    
    \item [\emph{Pooling layers}:] This layer reduces the  dimensionality of the set of convolved images. It applies some function (e.g. sum, mean, maximum) to a group of  spatially connected pixels and reduces the dimensions of such group. For example, it takes 2 consecutive pixels and converts them to the mean of both. Although we use it to handle the amount of data generated after the convolutions, it also regularises the model to avoid learning from non-generalisable noise and details in the training data (also known as overfitting). 
    \item [\emph{Fully connected layer}:] This layer is usually the last layer of the network. Its input is the linearised outcome of the previous ones (in our network, convolutions + poolings). It applies a linear transformation from the input to the output. The slope and bias of the linear transformation are the learning parameters. 
    \item [\emph{Batch normalisation layer}:] In this layer the network normalises the output of a previous activation layer. It subtracts the  mean and divides by the standard deviation. Batch normalisation helps to increase the stability of a neural network and avoids over-fitting problems.
\end{description}
After each convolution and fully connected layer there is an activation function that transforms the outcome. Activation functions are non-linear functions that map the outcome of a layer to the input of the following one. An example of an activation function is the Rectified Linear Unit (ReLu) \citep{Alexnet}, although we use a variation of this function called LeakyReLu \citep{LReLu}, with which we find better results.\\

Other terms that one needs to be familiar with are \textit{epoch} and \textit{batch}. An \textit{epoch} is an iteration over the complete training dataset. It is common practice to avoid feeding the network with all the training sample together. Instead, the training data is divided in groups of a certain size and each of these groups is called \textit{batch}. Feeding the network in batches helps it learn faster as in every iteration over a batch, it back-propagates updating all the weights. Then, instead of updating once per epoch, it updates as many times as there are batches. The amount of variation allowed per iteration is regulated by the \textit{learning rate}.

\section{Variable annulus}
\label{A2}
\LC{Currently, the PAUS photometry pipeline uses an annulus region to estimate the astronomical background behind a target source. The inner and outer radii are fixed at 30 and 45 pixels from the source, respectively. However, for each galaxy, one could adjust the annulus parameters to minimize the effect of the target galaxy flux falling inside the ring and background variations between the target and the annulus location.}\\

\LC{We can quantify the amount of extra light falling inside the annulus coming from the galaxy ($\Delta_{\rm F}$) and scattered light ($\Delta_{\rm B}$). For a flat background, $\Delta_{\rm B}$ should be independent of the annulus location. However, this term does depend on the annulus location when the background varies. On the other hand, $\Delta_{\rm F}$ is minimized by an annulus further away from the source and depends on the galaxy size and the PSF. We define 
\begin{equation}
\Gamma \equiv \frac{|\Delta_{\rm F} + \Delta_{\rm B}|}{\sigma_{\rm b}},
\label{Gamma}
\end{equation} where $\sigma_{\rm b}$ is the error on the background subtraction. $\Gamma$ measures the relative error on the background prediction due to scattered light and the source contribution. We use $\Gamma$ to study the  effect of a variable annulus by minimising the quantity as a function of annulus radius.}\\

\LC{We have tested this on simulations. The background images are simulated as in section \ref{sec:data}. The images also contain simulated galaxies with the same size (r$_{\rm 50}$) and PSF distribution as observed PAUS galaxies. Using simulations allows us to evaluate $\Delta_{\rm B}$ on the background simulations and $\Delta_{\rm F}$ on the galaxy simulations. This method cannot directly be applied to observed images, since it requires distinguishing between the galaxy flux and the background. However, it shows under which conditions the annulus approach degrades. Our simulations only contain Poisson noise and scattered light. We know that real PAUCam images have a more complicated noise pattern and therefore, we would also expect the distributions of $\Gamma_{\rm opt}$ to shift towards higher values.} \\

\begin{figure}
\includegraphics[width=.5\textwidth]{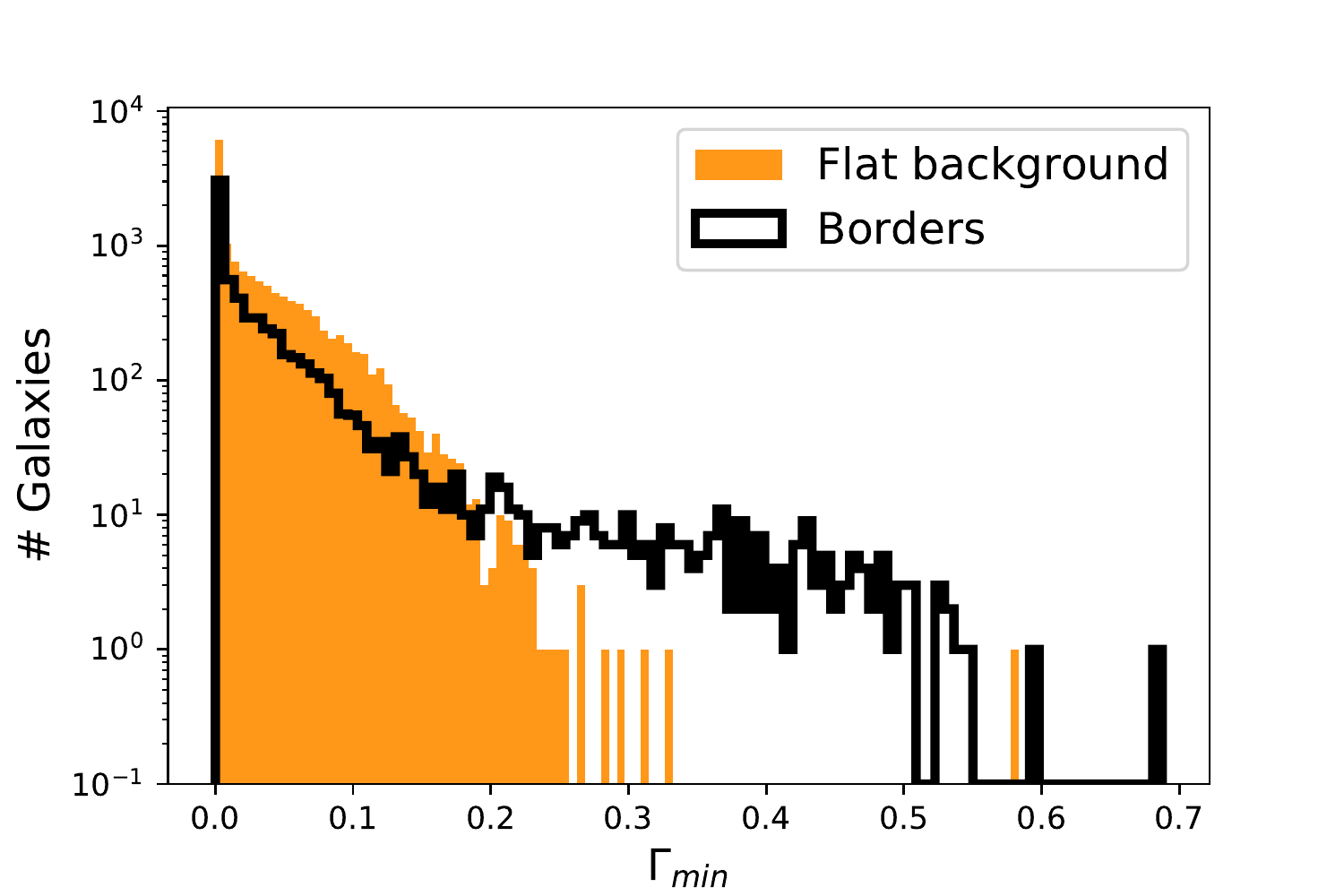}
\centering
\caption{The $\Gamma_{\rm min}$ distribution for galaxies in the border (scattered light affected, black) and in the center (flat background, filled orange) of the image.}  
\label{annulus_app}
\end{figure}

\LC{Figure \ref{annulus_app} shows the histogram of minimum $\Gamma$ measurements (Eq. \ref{Gamma}) for a set of simulated galaxies.
$\Gamma$ is evaluated for different annulus radii, moving the inner and outer radii between one and forty pixels from the source with fixed $r_{\rm out} - r_{\rm in}$ = 15 pixels. We split the results in galaxies in the center of the image (flat background, orange) and galaxies in the borders (scattered light, black). Both histograms show a large fraction of galaxies for which the annulus can predict the background very accurately (low gamma values). For galaxies in the center, we expect a flat background and therefore good results with the annulus method.}\\

\LC{There are also many galaxies (Fig. \ref{annulus_app}) on the border with accurate measurements. First of all, not all positions at the border are affected by scattered light. For those positions with scattered light, the annulus can be placed very close to the target galaxy if it is small ($r_{\rm 50} \approx $ 1 or 2 pixels). Then, the background variation from the target source to the annulus position would be small. Nevertheless, the distribution of border galaxies also shows a tail corresponding to galaxies for which the annulus estimation significantly degrades. In these cases, the optimal annulus is either too close to the source or capturing a strong background variation. If the background variation is very strong, the annulus will tend to get closer to the target source in order to minimize $\Delta_{\rm B}$. However, this is not possible for bright and large galaxies, since getting closer to the target increases $\Delta_{\rm F}$. The annulus method cannot be used to measure the background in large galaxies in varying background regions, since it strongly degrades. }

\end{appendices}

\bsp	
\label{lastpage}
\end{document}